\documentclass[fleqn,usenatbib]{mnras}

\usepackage{newtxtext,newtxmath}

\usepackage[T1]{fontenc}

\DeclareRobustCommand{\VAN}[3]{#2}
\let\VANthebibliography\thebibliography
\def\thebibliography{\DeclareRobustCommand{\VAN}[3]{##3}\VANthebibliography}

\usepackage{graphicx}	
\usepackage{amsmath}	
\usepackage{amssymb}	

\title[Helicity in the Large-Scale Galactic Magnetic Field]{Helicity in the Large-Scale Galactic Magnetic Field}

\author[J. L. West et al.]{
J. L. West,$^{1}$\thanks{E-mail: jennifer.west@dunlap.utoronto.ca}
R. N. Henriksen,$^{2}$
K. Ferri\`{e}re$^{3}$
A. Woodfinden$^{2}$
T. Jaffe,$^{4}$
\newauthor
B. M. Gaensler,$^{1}$
and J. A. Irwin$^{2}$
\\
$^{1}$Dunlap Institute for Astronomy and Astrophysics University of Toronto, Toronto, ON M5S 3H4, Canada\\
$^{2}$Queen's University, Kingston, ON K7L 3N6, Canada\\
$^{3}$Institut de Recherche en Astrophysique et Plan\'{e}tologie (IRAP), CNRS, Universit\'{e} de Toulouse, CNRS, 9 avenue du Colonel Roche, BP 44346,\\
31028 Toulouse Cedex 4, France\\
$^{4}$NASA Goddard Space Flight Center, Greenbelt, MD 20771, USA
}

\date{Accepted 28 September 2020. Received in original form 03 June 2020}

\pubyear{2020}

\begin{document}
\label{firstpage}
\pagerange{\pageref{firstpage}--\pageref{lastpage}}
\maketitle

\begin{abstract}
We search for observational signatures of magnetic helicity in data from all-sky radio polarization surveys of the Milky Way Galaxy. Such a detection would help confirm the dynamo origin of the field and may provide new observational constraints for its shape. We compare our observational results to simulated observations for both a simple helical field, and for a more complex field that comes from a solution to the dynamo equation. Our simulated observations show that the large-scale helicity of a magnetic field is reflected in the large-scale structure of the fractional polarization derived from the observed synchrotron radiation and Faraday depth of the diffuse Galactic synchrotron emission. Comparing the models with the observations provides evidence for the presence of a quadrupolar magnetic field with a vertical component that is pointing away from the observer in both hemispheres of the Milky Way Galaxy. Since there is no reason to believe that the Galactic magnetic field is unusual when compared to other galaxies, this result provides further support for the dynamo origin of large-scale magnetic fields in galaxies.\end{abstract}

\begin{keywords}
radio continuum: ISM -- magnetic fields -- cosmic rays -- polarization
\end{keywords}



\section{Introduction}
\label{sec:intro}
Magnetic fields are critical to the structure and the turbulent properties of the interstellar medium, to the star formation process, and to the acceleration, propagation and confinement of cosmic rays in galaxies \citep[e.g.,][]{2015ASSL..407..483H}. 
Observational features of galactic magnetic fields include a strength on the order of a $\mu G$ with magnetic field lines that generally follow the arms in face on spirals \citep{Beck:2001wn}. Observations of nearby, edge-on galaxies reveal apparent X-shaped field lines that extend into the halo \citep{2009Ap&SS.320...77B,Beck:2013gp,Krause:2015if, 2020A&A...639A.112K}. 

In recent years, we have learned a great deal about the magnetic field of the Milky Way Galaxy, both through modelling of the Galactic synchrotron radiation and Faraday rotation \citep{Page:2007ce,Sun:2008bw,Sun:2009jh,Sun:2010gk,  Jaffe:wl, Jansson:2012ep, Jansson:2012hl, Collaboration:2016eh, 2017A&A...600A..29T}, and through observations of the Faraday rotation of Galactic pulsars and background radio sources  \citep[e.g.,][]{Brown:2001jt, 2003ApJ...592L..29B,Pshirkov:2011gk,VanEck:2011hk, 2015A&A...575A.118O, 2019MNRAS.484.3646S, 2020A&A...633A.150H, 2020MNRAS.496.2836N}, yet we still do not have a clear picture of the 3D geometry of this field, nor do we fully understand the origin.  
The leading idea on how such ordered fields on galactic scales arise is through amplification of a weak seed field through dynamo action \citep{Beck:1996bn, Subramanian:2002uk}, however, conclusive evidence of this theory is yet to be achieved. 

One consequence of dynamo theory \citep{2015SSRv..188...59B, Subramanian:2002uk} is the presence of twisted, or helical, magnetic fields resulting from the Coriolis force, which produces a systematic rotation, always in the same sense (for expanding motions). Since all dynamo models will predict a field with a twist, confirming the presence of helicity in the magnetic field of a galaxy would strongly support a dynamo origin of the field. Additionally, a better understanding of the impact that helicity may have on the observed emission may help us devise new constraints on the geometry of galactic magnetic fields. In this paper, we set out to find observational evidence of helicity in the coherent, large-scale magnetic field of the Milky Way Galaxy.

Faraday rotation describes the rotation of the plane of polarization as an electromagnetic wave propagates through a magneto-ionic medium. Faraday depth, FD, is defined as the integral of the magneto-ionic medium along a line of sight, from a distance, $r=d$, to the observer at $r=0$,
\begin{equation}
\centering
\text{FD}(d) = 0.812\int\limits_0^d {{n_e}{B_\parallel }} dr {\rm{[rad~}}{{\rm{m}}^{\rm{-2}}}{\rm{]}},
\end{equation}
where $n_e$ [cm$^{-3}$] is the thermal electron density and ${B_\parallel }$ [$\mu$G] is the line-of-sight component of the magnetic field. Here ${B_\parallel>0}$ when pointed towards the observer. In this paper, we use $\text{FD}$ to mean the entire Faraday depth of the Galaxy, $\text{FD}(D)$, where $D$ is the distance to the edge of the Galaxy for a particular line of sight. 

Observationally, $\text{FD}$ is measured by looking at the dependence of the position angle of the polarization vector, $\chi$, as a function of the wavelength of observation squared, $\lambda^2$ [m$^2]$. If we measure the FD of a background source through a Faraday-rotating medium, then we call this the Faraday rotation measure, $\text{RM}$, where
\begin{equation}
\centering
\chi_\text{obs}=\chi_\text{src}+\text{RM}\lambda^2.
\end{equation}
The amount of rotation (i.e., the difference between $\chi_\text{obs}$ and $\chi_\text{src}$) is greater at longer wavelengths since the rotation is $\propto\lambda^2$.

While the degree of Faraday rotation depends on the line-of-sight component of the magnetic field, the amount of synchrotron radiation depends on the perpendicular component of the magnetic field (i.e., in the plane of the sky), meaning these observations combined could theoretically constrain the 3D magnetic field. However, since the magnetic field is observed in projection, and since Faraday rotation, as well as geometric effects (e.g., magnetic field reversals, turbulence, beam depolarization, etc.) can lead to cancellation of the polarization vectors (depolarization), the analysis is complex.

Magnetic helicity, $H_m$, is a property of a magnetic field, $\bf{B}$, that describes the amount of coil or twist that is present in the field. It is defined as $H_m=\langle {\bf{A}} \cdot {\bf{B}} \rangle$, where $\bf{A}$ is the vector potential and $\bf{B}=\nabla  \times \bf{A}$. However, there are an infinity of possible vector potentials that can satisfy this equation, which means that $H_m$ is not a uniquely defined quantity. The integral over a volume is unique only if the normal component of the magnetic field vanishes on the surface. Observationally, it is a better choice to consider the current helicity, $H_j$, since it is uniquely defined for a given $\bf{B}$. Current helicity is defined by the volume average of $ \bf{j} \cdot \bf{B}$, i.e.,
\begin{equation}
\label{eqn:helicity}
\centering
H_j= \langle {\bf{j}} \cdot {\bf{B}}  \rangle, 
\end{equation}
where $\bf{j}=\nabla  \times \bf{B}$ is the current density  \citep[e.g.,][]{1990SoPh..125..219S}. Both magnetic helicity and current helicity are measures of the amount of coil or twist in the magnetic field. Although there is no general equation that relates magnetic helicity and current helicity, at the relatively large scales considered in this paper, both helicities should have the same sign.

The sense of this twist has a handedness that can be right-handed (positive helicity) or left-handed (negative helicity). \citet{Junklewitz:2011ha} develop a methodology and \citet{Oppermann:2011fc} use this method to attempt to detect helicity in the turbulent component of the Galactic magnetic field, but its presence can not be confirmed by these observations. More recently, \citet{2020arXiv200314178B} use observations of B-mode polarization in Wilkinson Microwave Anisotropy Probe (WMAP) data to demonstrate broad agreement with a model, which is suggestive of opposite handedness in the North and South Galactic hemispheres.

\citet{Volegova:2010go} present a study on detecting helicity in a purely turbulent field (i.e., not specific to the magnetic field of a galaxy or using a dynamo field). In this study, they predict a relationship between polarized fraction\footnote{Polarized fraction is also referred to as the degree of polarization.}, $P/I$, where $P$ is the linearly polarized flux density and $I$ is the total Stokes I flux density, and the observed $\text{RM}$, which depends on the helicity of the field. When there is no helicity, the rotation measure distribution is symmetric and the cross-correlation coefficient, $C$, between polarized fraction and rotation measure is consistent with zero. When $H_m>0$, the distribution is tilted towards negative $\text{RM}$ for large polarized fractions, which leads to $C>0$. In contrast, when $H_m<0$, there are more negative $\text{RM}$ for small polarized fractions and more positive $\text{RM}$ at large polarized fractions, and hence  $C<0$. Work by \citet{Brandenburg:2014gp} further developed this idea.

The explanation for this correlation is in the way that Faraday rotation interacts with a helical field. Faraday rotation rotates the plane of polarization in a right-handed sense about the magnetic field, i.e., if the magnetic field is pointing towards the observer ($\text{FD}>0$) then the polarization vectors are rotated counter-clockwise as they propagate towards the observer \citep[e.g.,][]{2018arXiv180607391R}. Therefore Faraday rotation either causes a ``winding'' of the orientation of the polarized electric field vector, consistent with the direction of Faraday rotation (causing greater depolarization), or ``unwinding'', which is rotation of the polarized electric field vector opposite to the direction of Faraday rotation (causing lesser depolarization), depending on the sign of the $\text{FD}$ and the handedness of the helicity \citep[for further explanation see][]{Faraday-rotation-helicity}.

For example, if the magnetic field has left-handed helicity ($H_m<0$) and points towards the observer ($\text{FD}>0$), then Faraday rotation causes an ``unwinding'', resulting in decreased depolarization (larger polarized fraction). This causes a bias in the observed polarized fraction for a particular sign of FD, leading to a correlation between the two quantities. The theoretical maximum polarized fraction is $\approx75\%$. Helicity cannot increase this value, but it can act to decrease it through increasing the depolarization effect. 

\citet{Volegova:2010go} find that for simulations with $\lambda<6$~cm, i.e., wavelengths for which the amount of Faraday rotation is small, $C$ is almost equal to zero. They find the strongest correlation to be $C \approx 0.4$ for $\lambda=15$~cm ($\nu=2$~GHz), where the degree of Faraday rotation is $\gtrsim6$ times larger.

\begin{figure*}
\centering 
\begin{minipage}{8.8cm}
\includegraphics[width=8.8cm]{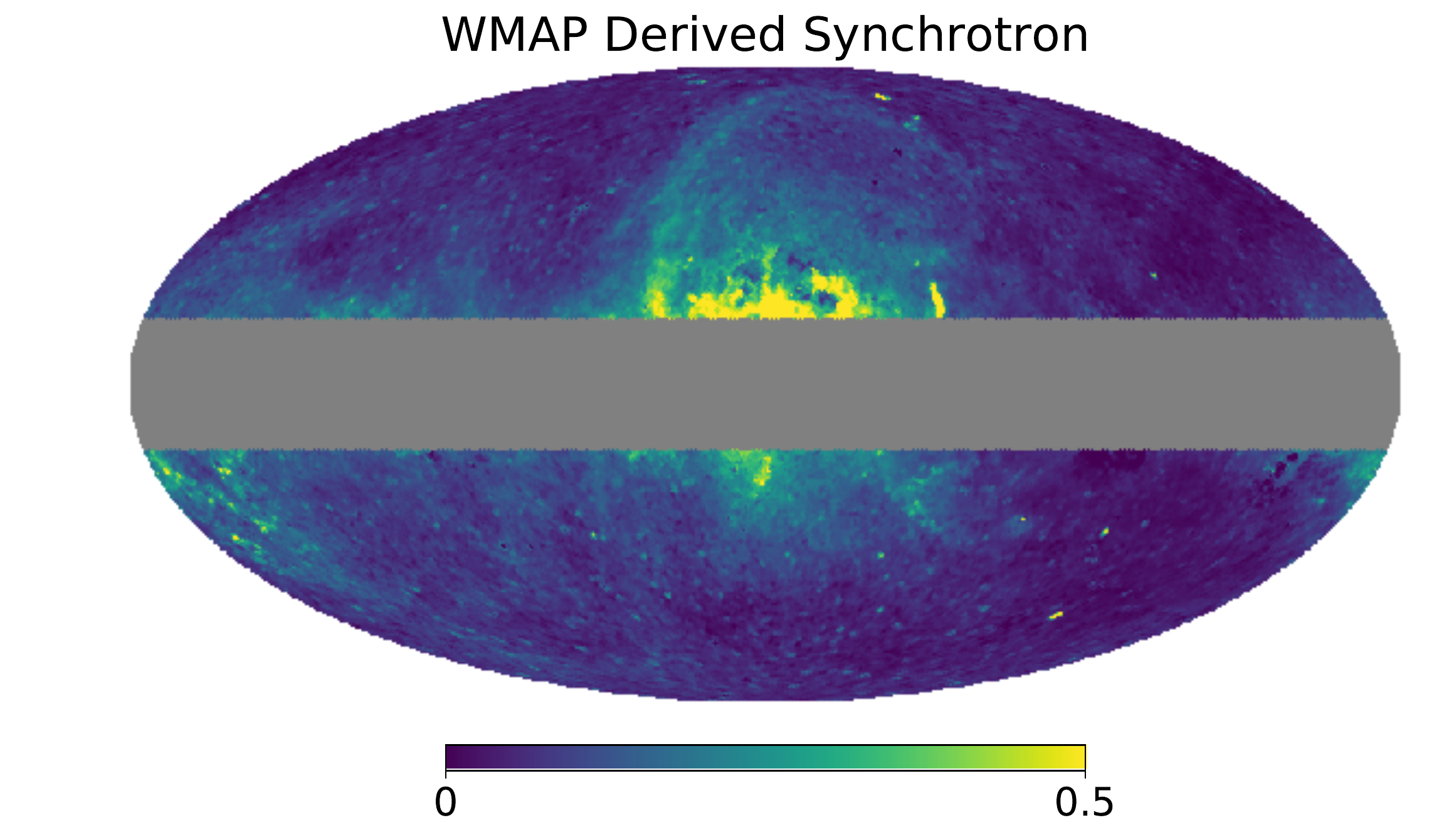}

\end{minipage}
\hfill
\begin{minipage}{8.8cm}
\includegraphics[width=8.8cm]{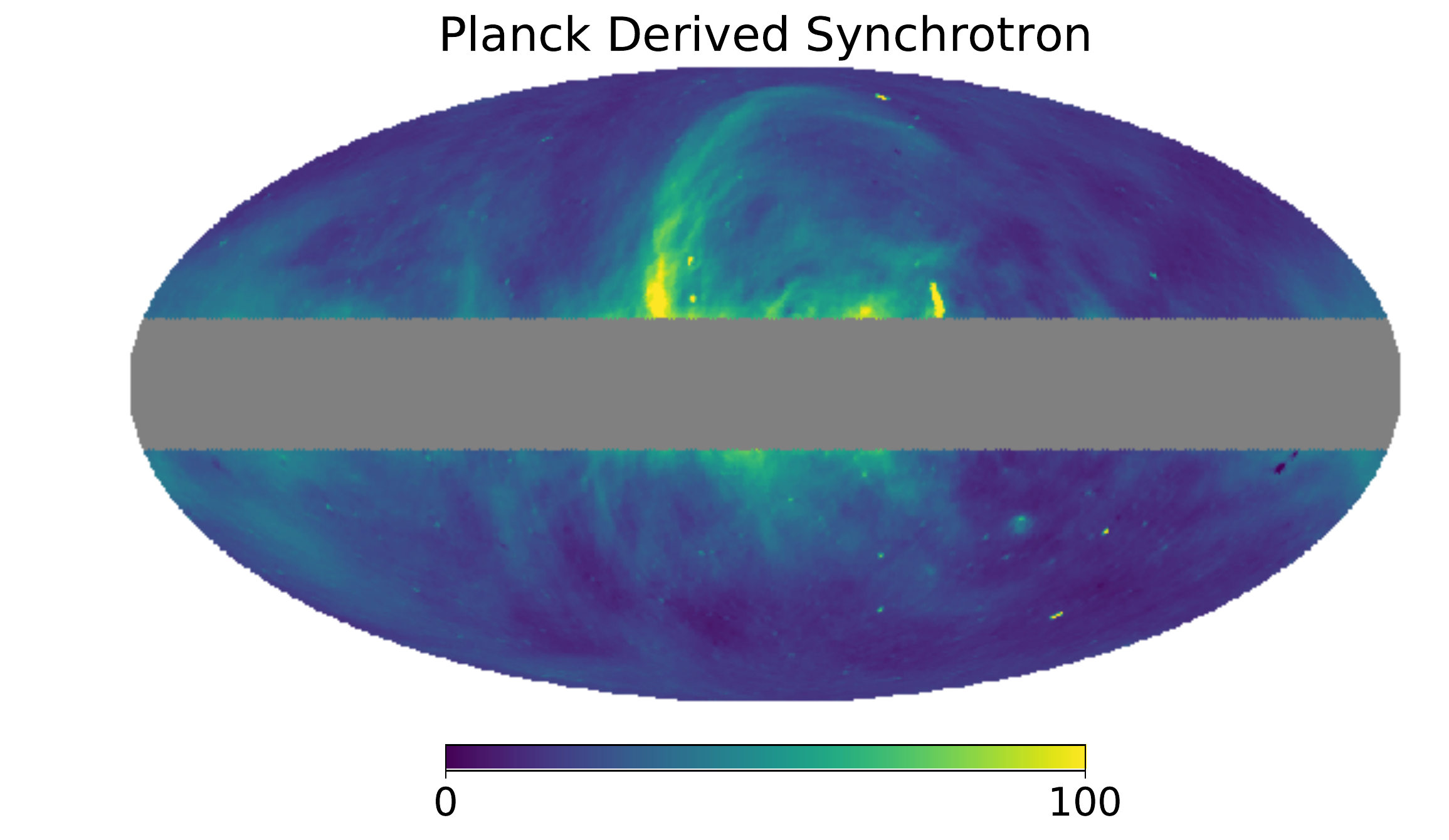}

\end{minipage}
\begin{minipage}{8.8cm}
\includegraphics[width=8.8cm]{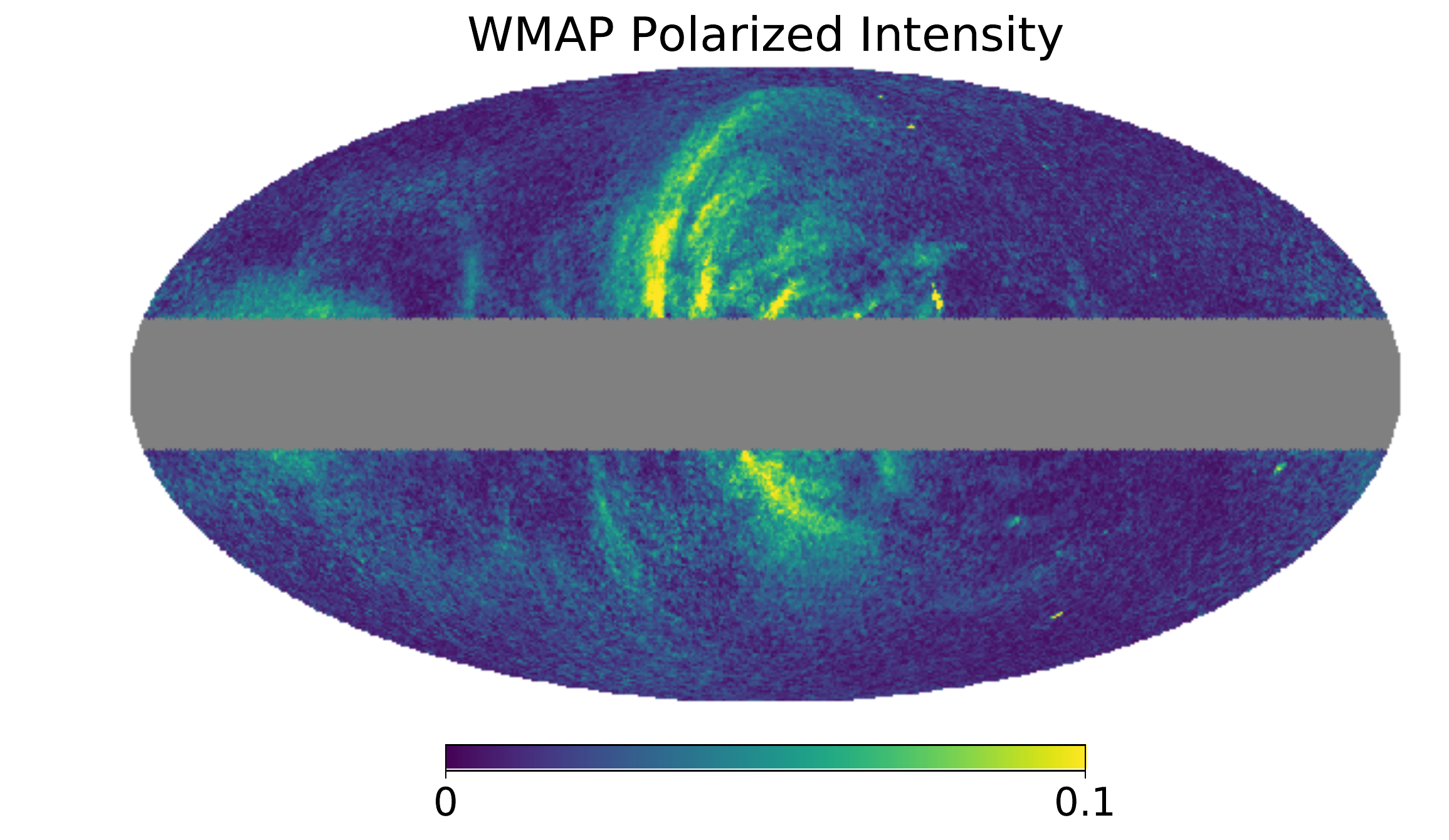}
\end{minipage}
\hfill
\begin{minipage}{8.8cm}
\includegraphics[width=8.8cm]{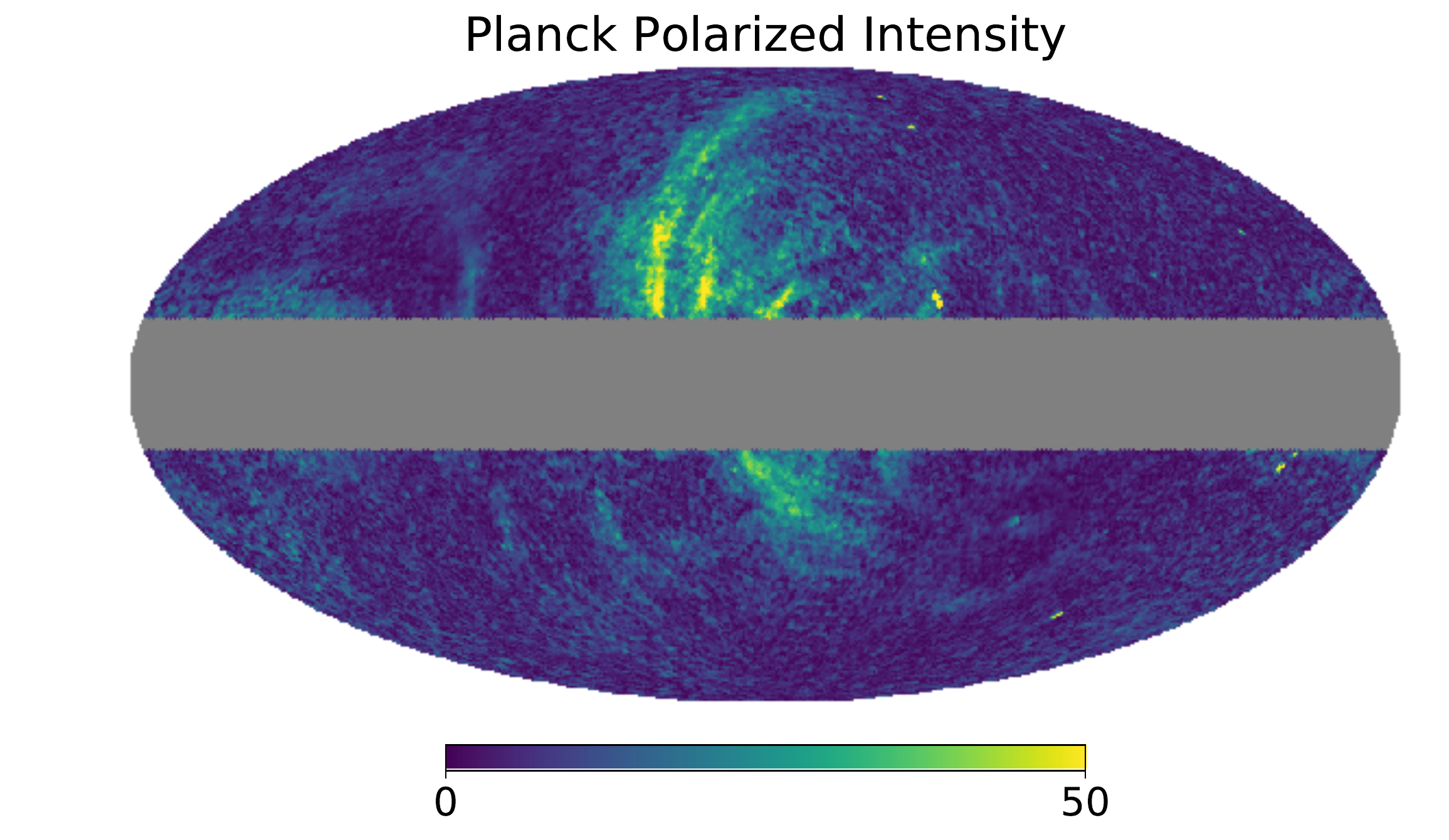}
\end{minipage}
\hfill
\begin{minipage}{8.8cm}
\includegraphics[width=8.8cm]{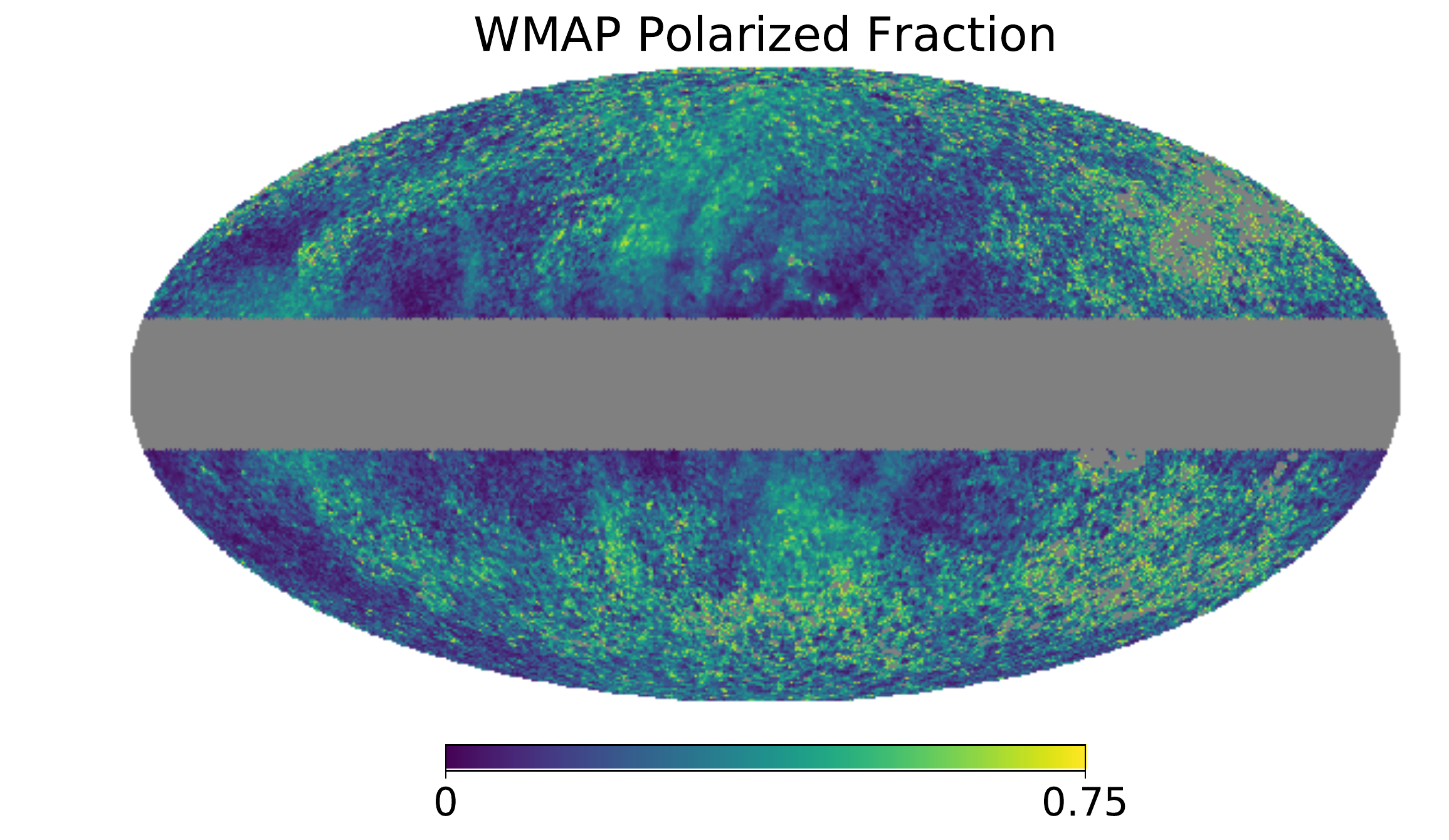}
\end{minipage}
\hfill
\begin{minipage}{8.8cm}
\includegraphics[width=8.8cm]{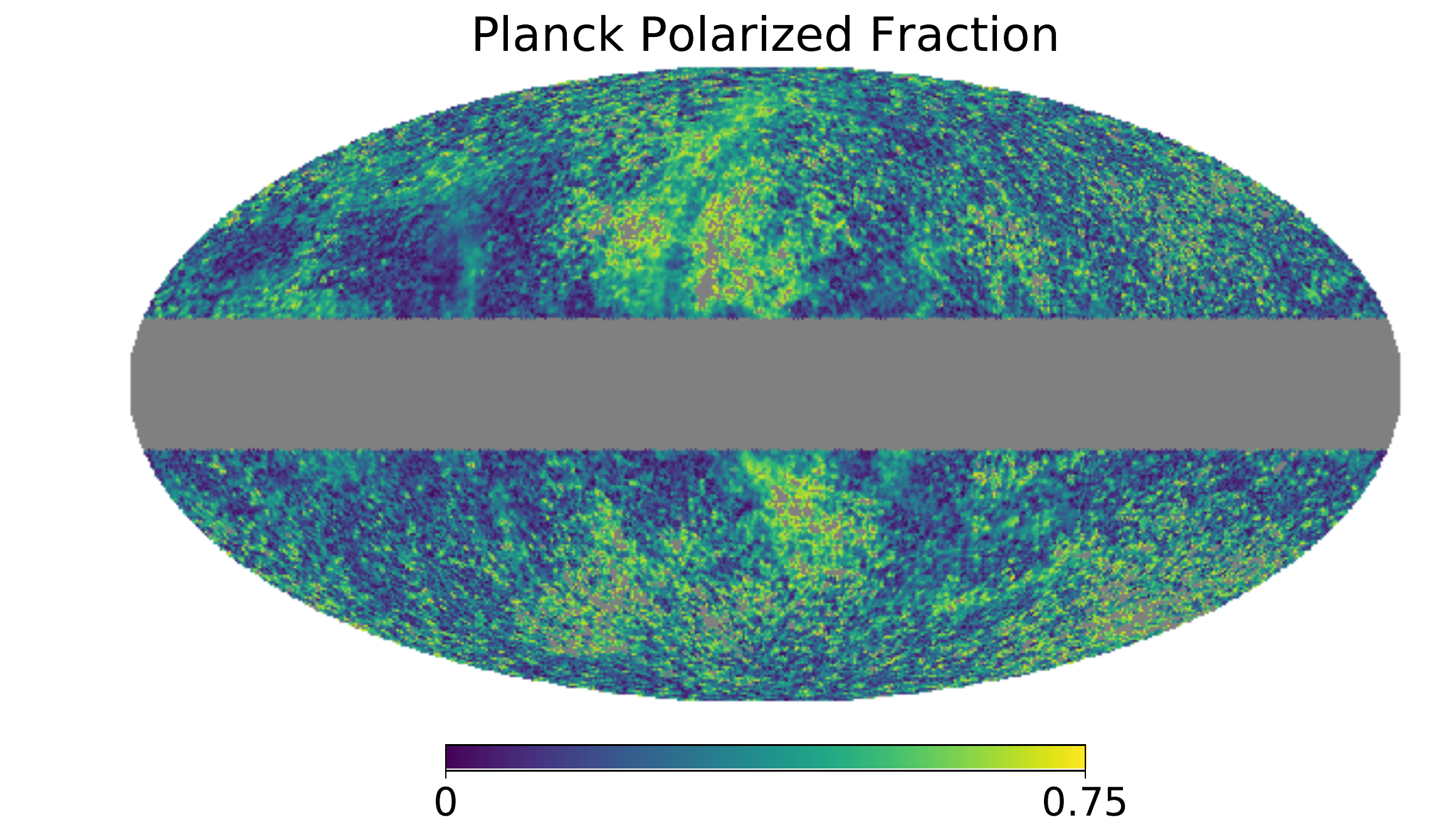}
\end{minipage}
\hfill
\begin{minipage}{8.8cm}
\includegraphics[width=8.8cm]{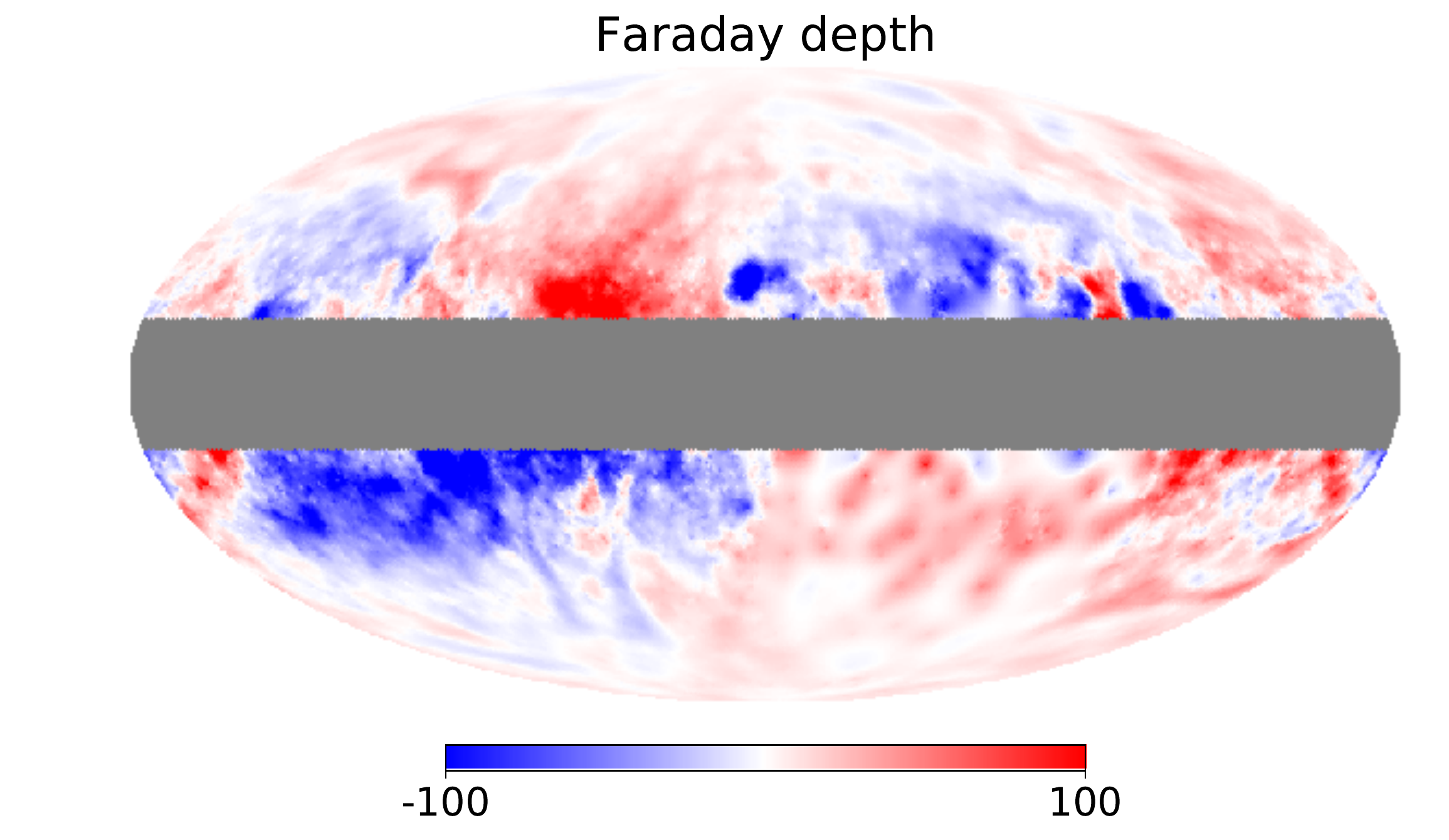}
\end{minipage}

\caption{ \label{fig:wmap}Left column: Derived synchrotron total intensity (top row) and polarized intensity (second row) (in units of mK) for the 23~GHz WMAP data shown with the polarized fraction map (third row). Right column: Derived synchrotron total intensity (top row) and polarized intensity (second row) (in units of $\mu$K) for the 30~GHz Planck data shown with the polarized fraction map (third row). Bottom row: The Galactic FD map of \citet{2015A&A...575A.118O}, in units of rad~m$^{-2}$, is included for ease of comparison. In all cases, the region for $|b|<15^\circ$ has been masked out as it is not included in these analyses. The orientation of all maps is the same, shown in Galactic coordinates in Mollweide projection, with the Galactic centre at the centre of the map and the Galactic plane oriented horizontally through the centre of each map. Masked values (shown as grey) are not included in the computations.}
\end{figure*}

\begin{figure}
\centering \includegraphics[width=8.3cm]{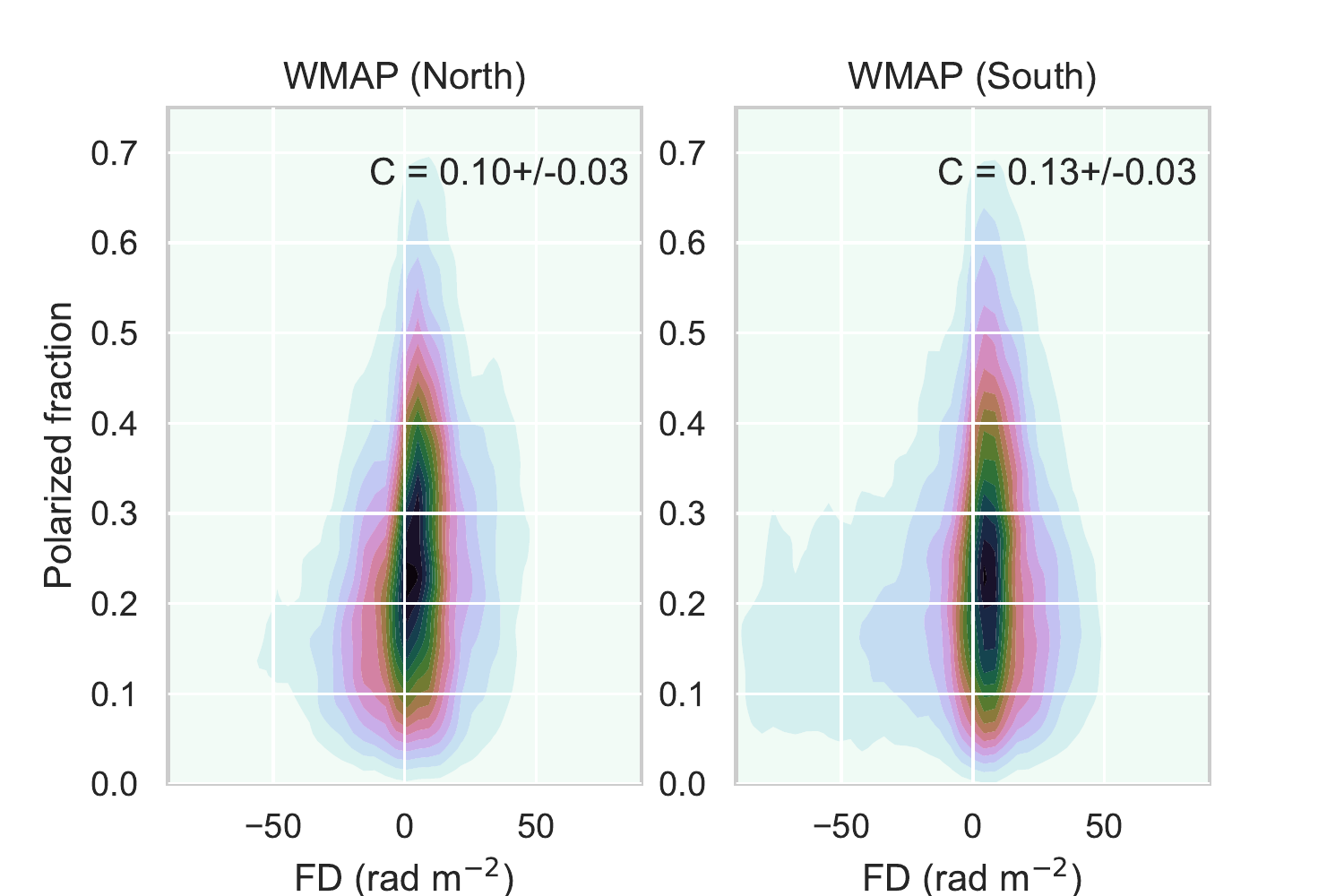}
\centering \includegraphics[width=8.3cm]{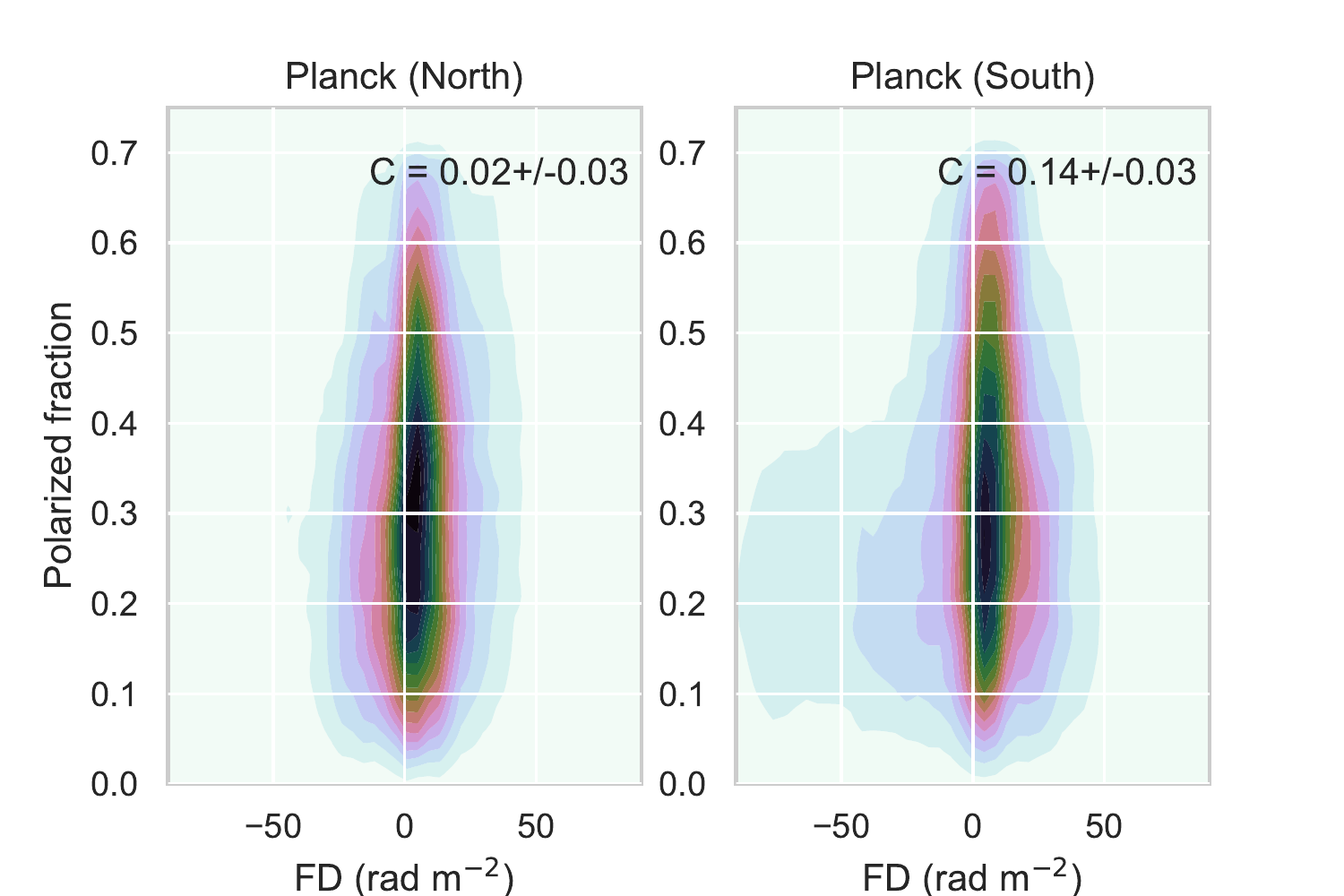}
\begin{scriptsize}\caption{\label{fig:wmap-density}2D-histogram of polarized fraction and FD for the Northern (left) and Southern (right) Galactic hemispheres ($|b|>15^\circ$) in the WMAP 23~GHz data (top) and Planck 30~GHz data (bottom). }
\end{scriptsize}

\end{figure}

In the case of the simulations described above, the simulation box contains pure turbulence with no coherent, large-scale pattern in the magnetic field. However, we know that the magnetic field of a galaxy does have a coherent pattern. And in addition to helicity that may be present in the turbulent field, the mean-field dynamo also has helicity present in the coherent, large-scale field.

The alpha effect of the mean-field dynamo is a process that generates one component of the large-scale magnetic field from another (e.g., $B_r$ from $B_\phi$). This is illustrated by \citet{1970ApJ...162..665P}, who shows how a rotating turbulent cell rising from an azimuthal magnetic field can produce a magnetic loop. Coalescence of many such loops leads to a large-scale poloidal field. Thus, magnetic helicity injected at turbulent scales is transferred to the largest Galactic scales. This transfer occurs on the long time scale  \citep[on the order of ten rotation periods according to ][]{2018JPlPh..84d7304B} of the mean-field dynamo, and preserves the sign of the helicity. Comparison of the observed helicity with the helicity predicted by galactic dynamo models could provide constraints on the alpha effect, and hence on the mean-field amplification rate.

If there is a coherent, large-scale Galactic field with helicity (i.e., twist) then the linear polarization pseudovector may rotate along the line of sight so as to give a small or zero net polarization, even when observed at frequencies where there is negligible Faraday rotation. And since $\text{FD}$ is a physical quantity, which depends only on the column density of thermal electrons and the configuration of the magnetic field along the line of sight (i.e., $\text{FD}$ is not a function of observation frequency), $\text{FD}$ should be correlated to some degree with the net linear polarized fraction for a helical field. This should be true even in high-frequency data such as the WMAP 23~GHz (1.5~cm) and Planck 30~GHz (1~cm) data. For reference, there is nearly 1000 times more Faraday rotation at 1~GHz than at 30~GHz. Using a Galactic FD of 100~rad~m$^{-2}$, we would expect $\Delta\chi\simeq0.6^\circ$ at 30~GHz, but $\Delta\chi\simeq515^\circ$ (more than 1 full rotation of the polarization pseudovector) at 1~GHz. This amount of rotation applies through the entire Galactic path length for a particular line-of-sight, which is typically on the order of a few~kpc, depending on the Galactic latitude.

In this paper, we investigate whether such correlations are possible to detect in a coherent, Galactic-scale magnetic field, and whether these correlations are consistent with signatures of helicity. 
In Sec. \ref{sec:data} we present our search for such a correlation in the Milky Way Galaxy by cross-correlating large-scale polarized emission from several large-scale surveys and comparing to measurements of Galactic FD. In Sec.~\ref{sec:modelling}, we present our modelling, which uses the Hammurabi code\footnote{\url{http://sourceforge.net/projects/hammurabicode/}} \citep{Waelkens:2009bn} (described in Sec.~\ref{sec:hammurabi}) to generate synthetic polarized fraction and FD maps using a model magnetic field. We first use a simple, toy model of a singly helical large-scale field to investigate whether the case of short-wavelength (i.e., high-frequency) observations, with negligible Faraday rotation, can still cause a correlation between the Faraday rotation measure and the polarized fraction (Sec.~\ref{sec:toymodels}). We investigate the trends in the correlation as we change the observation frequency or introduce turbulence. We then then use a more physically motivated magnetic field model, which comes from a solution to the dynamo equation, to further investigate whether this effect could be detectable in observations (Sec.~\ref{sec:dynamo}). The discussion and conclusions are presented in Sec. ~\ref{sec:discussion} and \ref{sec:conclusions}, respectively.

\section{Data}
\label{sec:data}

We look for signatures of helicity by a cross-correlation analysis to compare FD to polarized fraction data, as described by \cite{Volegova:2010go}. 

A widely-used Galactic FD map (see Fig.~\ref{fig:wmap}) is that derived by \citet{2015A&A...575A.118O}\footnote{\url{https://wwwmpa.mpa-garching.mpg.de/ift/faraday/2014/index.html}}. This map is produced by observations of Faraday rotation of extragalactic sources that probe the entirety of the path through the Galaxy to the observer. The authors use a careful reconstruction technique to separate the Galactic foreground contribution from the extragalactic component.

We use their HEALPix\footnote{\url{http://healpix.sourceforge.net}} map for this analysis, which is provided at a resolution of $N_{\text{side}}=128$ (pixel size is $0.45^{\circ}$). The HEALPix pixelization scheme \citep{Gorski:2005ku} is designed such that each pixel represents an equal angular area on the sky. It is ideal for this study since other sky projections may introduce biases or  correlations resulting from projection effects. 

Polarized fraction is computed using $P=\sqrt{{Q^2}+{U^2}}$, where $Q$ and $U$, are the linear polarization Stokes parameters. It is a difficult quantity to determine well due to many inherent uncertainties:
\begin{enumerate}
\item In regions of low signal to noise in Stokes I, the fractional polarization values will be very uncertain. 
\item  The absolute zero point of many surveys is difficult to calibrate and often has some reasonably large uncertainty. 
\item Depolarization becomes significant at lower frequencies ($<$2 GHz), which means that the polarized fraction will be suppressed over much of the sky. 
\item At higher frequencies, such as from the WMAP (23 GHz) and Planck (30~GHz) satellites, data suffer from uncertainty due to increasing contributions from thermal dust and free-free emission, making it necessary to estimate the contribution from the non-thermal synchrotron component in these maps. These derived synchrotron maps are especially uncertain for the region in and around the Galactic plane \citep{Collaboration:2016eh}. These systematic errors are much harder to quantify, and are likely to be more important that statistical uncertainties in these data.
\item Since $P$ has a Rician, rather than Gaussian, noise distribution, with a non-zero mean noise value, there is a noise bias present.
\end{enumerate}
In order to mitigate these uncertainties, we perform this analysis using two independent data-sets (details are provided in Sec.~\ref{sec:wmap} and Sec.~\ref{sec:planck}). Any signal common to both is likely real.

We also exclude the Galactic plane ($|b|\leqslant 15^\circ$), and additionally mask unphysical polarized fraction values (i.e., $P/I < 0$ and $P/I > 0.7$) for all measurements. See Sec.~\ref{sec:dataresults} for a discussion on how we test the impact of the noise bias.

In each case we calculate the Pearson correlation coefficient\footnote{calculated in this work using Python 2.7 and numpy.corrcoef \citep{2020SciPy-NMeth}}, $C$, for the Northern and Southern Galactic hemisphere separately. Each HEALPix map with $N_{\text{side}}=128$ has 196,608 resolution elements, which is reduced to 72,960 elements when selecting only the high-latitude elements for a single hemisphere. We use bootstrapping and draw 1000 elements from this sample to compute a value of $C$. We then find the mean and standard deviation over 1000 iterations (each iteration using 1000 samples) to determine the average measurement of $C$ with a 1$\sigma$ uncertainty.

\subsection{WMAP 23 GHz data}
\label{sec:wmap}
For the Stokes $I$ map, we use the foreground K-band (23~GHz) synchrotron component derived using the maximum entropy method \citep{2013ApJS..208...20B}. We use the full nine years of data for Stokes $Q$ and $U$ \citep{2013ApJS..208...20B}. The WMAP 23~GHz maps have a native resolution of 53$'$. We resample with $N_{\text{side}}=128$ to match the FD map.

These WMAP data, along with the \citet{2015A&A...575A.118O} FD map and derived polarized fraction maps, are shown in Fig.~\ref{fig:wmap}. The 2D-histogram showing the cross-correlation coefficient of polarized fraction vs $\text{FD}$ is shown in Fig.~\ref{fig:wmap-density}. 

It can be clearly seen in this plot that the distribution of polarized fraction vs $\text{FD}$ is skewed. It is particularly clear in the South that for low polarized fractions ($<0.3$), there are more points with $\text{FD}<0$ than $\text{FD}>0$. This is also true in the North, but to a lesser degree. In addition, in the North one can also see a slight excess of higher polarized fractions ($>0.3$) where $\text{FD}>0$.

\subsection{Planck 30 GHz data}
\label{sec:planck}
For the Planck data, the foreground (30 GHz) synchrotron component is derived using the 408 MHz map, originally from \citet{1982A&AS...47....1H}, which has been scaled to 30 GHz using a constant synchrotron spectrum corresponding to a cosmic-ray electron (CRE) power law index, $p=-3.1$ \citep{2016A&A...594A..10P}.  We also use the Planck Stokes $Q$ and $U$ maps at 30~GHz to derive a polarized fraction map. The Planck 30~GHz data has a resolution of 33$'$. We smooth this to 60$'$ and resample with $N_{\text{side}}=128$ to match the FD map.

These Planck data along with the the \citet{2015A&A...575A.118O} FD map and derived polarized fraction maps are shown in Fig.~\ref{fig:wmap}. The 2D-histogram showing the cross-correlation coefficient of polarized fraction vs FD is shown in Fig.~\ref{fig:wmap-density}.

Similar to the WMAP data, it is clear that in the South for low polarized fractions ($<0.4$), there are more points with $\text{FD}<0$ than $\text{FD}>0$. The distribution in the North in this case is quite symmetric, and thus $C$ is consistent with zero in this case.


\begin{table}
\centering
\begin{tabular}{|c|c|c|}
\hline 

& Northern hemisphere & Southern hemisphere\tabularnewline

\hline 
\hline 

WMAP 23~GHz & $0.10\pm0.03$ & $0.13\pm0.03$  \tabularnewline

Planck 30~GHz & $0.02\pm0.03$ & $0.14\pm0.03$  \tabularnewline

\hline 

\end{tabular}
\caption{Cross-correlation coefficient, $C$, between $\text{FD}$ and polarized fraction.}
\label{tab:correlations}
\end{table}


\begin{table}
\centering
\begin{tabular}{|c|c|c|}
\hline 

& Northern hemisphere & Southern hemisphere\tabularnewline

\hline 
\hline 

WMAP 23~GHz & $-4.1 \times10^{-5} \pm0.006$ & $-5.5 \times10^{-5} \pm0.006$  \tabularnewline

Planck 30~GHz & $-5.5 \times10^{-5} \pm0.004$ & $1.0 \times10^{-4} \pm0.004$  \tabularnewline

\hline 

\end{tabular}
\caption{Cross-correlation coefficient, $C$, between $\text{FD}$ and randomized polarized fractions.}
\label{tab:correlations-random}
\end{table}

\subsection{Results from the observations} 
\label{sec:dataresults}
All values of $C$ are summarized in Table~\ref{tab:correlations}. For both WMAP and Planck data we measure a small but significant positive correlation in the Southern Galactic hemisphere. The results for the Northern Galactic hemisphere are less clear. Planck does not measure any significant correlation for the North while WMAP does, although with less significance than the detection in the South.

In order to check the robustness of the detection and test whether the correlations are just a function of FD, independent of the polarized fraction, we randomly shuffle the polarized fraction values in the array and recalculate $C$. We find that $C$ is consistent with zero for all cases. The mean and standard deviation of 1000 iterations of 1000 random samples are summarized in Table~\ref{tab:correlations-random}.

The FD map from \citet{2015A&A...575A.118O} also includes a map of the uncertainty, dFD. We quantify the impact of this uncertainty on $C$ by taking the maximum value of dFD and randomly adding or subtracting this to the FD map and recomputing $C$. With this test we determine that the uncertainty introduced here is smaller than that derived from the bootstrapping ($\sim\pm0.02$ from the dFD map compared with $\pm0.03$ from bootstrapping). We note that the dFD map is derived in a complicated way that includes some contribution from Galactic variance that we are also sampling in the bootstrapping. Thus, these uncertainty measurements are not independent. We therefore quote the uncertainty from the bootstrap method since this is the larger of the two. 

We also note that the reconstruction used to make the FD map  suffers from sparsely sampled measurements for Southern declinations. This results in higher uncertainties for this region, which in Galactic coordinates is located at high longitudes in the Southern Galactic hemisphere, i.e., to the right of the Galactic centre ($240^\circ<l<360^\circ$) and on the lower side ($b<0$), where the FD is mostly positive (bottom panel of in Fig.~\ref{fig:wmap}). However, this is unlikely to impact our conclusion since the asymmetry that can be seen in the 2D-histograms shown in Fig.~\ref{fig:wmap-density} are skewed predominantly by values of $\text{FD}<0$ rather than $\text{FD}>0$.

 We test the impact of the noise bias by exploiting the fact that both the Planck and WMAP data were observed over a number of years, and independent maps of Stokes $Q$ and $U$ with different noise were produced. One method to correct for this bias is to multiply these maps together when making the $P$ map. I.e., if $Q_{1}$ [$U_{1}$] is the Stokes~$Q$ [$U$] map made for the first half of the mission and  $Q_{2}$ [$U_{2}$] is the Stokes~$Q$ [$U$] map made for the second half of the mission, then we can use $P= \sqrt{  Q_{1}Q_{2}+U_{1}U_{2} }$ to make a bias corrected $P$ map. We make these bias corrected $P$ maps and use this to make a new polarized fraction map and recalculate $C$. We find the result remains consistent with the values shown in Table~\ref{tab:correlations}.

\section{Models}
\label{sec:modelling}
In the previous section, we presented a detection of a correlation between FD and polarized fraction in observations. In this section we use toy models of coherent Galactic-scale magnetic fields to test the idea that these correlations are due to helicity in the large-scale field of the Milky Way Galaxy.

The models we use are all quite simple and none are intended to truly represent the Galactic magnetic field in detail. Rather the goal is to use these models to investigate trends in how such a correlation may vary as a function of frequency for magnetic field models that have helicity of known handedness. We also investigate how these trends are impacted when we include a random component, which is added to the coherent component of the field.

We use a coordinate system that defines the plane of the model galaxy to be parallel to the $xy$-plane, with the origin at its galactic centre. The $z$-axis is perpendicular to the plane, with $z>0$ towards the Northern hemisphere. Despite being simplified models, we still use Galactic-like properties in several respects:

\begin{enumerate}
    \item The models use a grid with a bounding box physical size of 40~kpc $\times$ 40~kpc $\times$ 10~kpc (in the $x$-, $y$-, and $z$-coordinates, respectively) in order to use a scale similar to that of the Galaxy. 
    \item We place an observer inside of this field, at a position analogous to the Sun's position in the Galaxy, i.e., $(x,y,z)=(-8.5,0,0)$~kpc. In Sec.~\ref{sec:dynamo} we use the electron density model of \citet{2017ApJ...835...29Y}, which assumes a slightly different Solar position \citep[$x=-8.3$~kpc from ][]{2011AN....332..461B}. However this has no impact on our results as we remove all nearby structures (see discussion in Sec.~\ref{sec:dynamo}) and there is no structure on the scale of this small, 200~pc difference in Solar position.
    \item The strength of the coherent magnetic field is on the order of $\sim1~\mu$G, which is similar to the Galactic field. The strength of the coherent magnetic field in the plane of the Milky Way is thought to be about 5~$\mu$G \citep{2015ASSL..407..483H}. The strength of the halo field is less well known, but thought to be $\approx1$-2.5~$\mu$G \citep{2015ASSL..407..483H}. Our analysis excludes the Galactic plane, and thus focuses more on the Galactic halo where the field is weaker.
    \item We add a random component of varying strength, up to $6~\mu$G, which is of the order expected from turbulence in the Galaxy \citep{2015ASSL..407..483H}. A relevant related quantity is the ratio of the strength of the random magnetic field component to the regular component. This quantity has several estimated values from different works with estimates of this ratio ranging from $<1$ to $\sim2$ \citep[][and references therein]{2015ASSL..407..483H}, depending on the particular measurement and the location in the Galaxy (i.e., disc vs halo). These factors motivate our decision to test several different values of the random component strength and thus different values for the random to regular field strength ratio.
    
\end{enumerate}

We integrate the coherent field alone using a low-resolution Healpix $N_\text{side}=64$, corresponding to roughly $1^\circ$ pixels.  This angular resolution corresponds to a physical distance that varies along the LOS and is roughly 200~pc at a distance of 10~kpc.  These models have no small-scale structure.  

In the cases where we introduce a random magnetic field component, a higher resolution model and integration are necessary to resolve some smaller-scale structure and its averaging effects.  We use a Cartesian grid of dimension 512~pixels $\times$ 512~pixels $\times$ 128~pixels (physical width $\simeq 8$~pc per cell) to define the fields (see below) and integrate with a Healpix grid that varies from $N_\text{side}=32$ to $N_\text{side}=512$  as a function of distance to maintain a width of roughly 50~pc on average (50~pc pixels correspond to $\sim3^\circ$ at a distance of 1~kpc and to $\sim0.3^\circ$ at 10~kpc). \citep[See][for how the Hammurabi code handles this.]{Waelkens:2009bn}

\subsection{Method}
\label{sec:hammurabi}
The Hammurabi code was created to model the large-scale structure of the Galactic magnetic field. It models the synchrotron emission and Faraday rotation given an
input 3D magnetic field, thermal electron distribution, and CRE distribution, for an observer that is embedded in the observed volume. There is no absorption included in these models, so the assumption is that the medium is optically thin.

Hammurabi calculates the simulated Stokes $I$, and the Stokes $Q$ and $U$ parameters relevant to the linear polarization, which are expressed as:
\begin{equation}
\label{eqn:hammurabi}
\begin{aligned}
& {I_{i}}={C_{I}}B_{i,\bot}^{\left({1-p}\right)/2}{\nu^{\left({1+p}\right)/2}}\Delta r
\\
& {P_{i}}={C_{P}}B_{i,\bot}^{\left({1-p}\right)/2}{\nu^{\left({1+p}\right)/2}}\Delta r
\\
& \Delta \text{FD}_{i}=0.812{n_{e}}{B_{i,\parallel}}\Delta r
\\
& \text{FD}_i = \sum\limits _{j=1}^{j=i}{{\Delta \text{FD}_{j}}}
\\
& \chi_i = \chi_{i,0} + \text{FD}_i \, \lambda^2
\\
& {Q_{i}}={P_{i}}\cos\left({2{\chi_{i}}}\right)
\\
& {U_{i}}={P_{i}}\sin\left({2{\chi_{i}}}\right).
\end{aligned}
\end{equation}
Here, $i$ corresponds to the $i$-th volume element along some line of sight, $p$ is the CRE 
power law spectral index (where {\bf $dN/dE$ $\sim$ $E^p$}), $C_{I}$ and $C_{P}$ are factors that are
dependent on $p$ \citep[see][]{Waelkens:2009bn,1979rpa..book.....R}. $\text{FD}_i$ is the FD of the i-th element and $\Delta \text{FD}_i$ is its Faraday thickness. The intrinsic polarization angle of the $i$-th element, $\chi_{i,0}$, is defined as the inclination angle of the plane-of-sky component of the magnetic field, $B_{i,\perp}$, with respect to north (in the frame of Galactic coordinates), rotated by $90^\circ$. The polarization angle that would be observed from emission at the $i$-th element, $\chi_i$, is given by $\chi_{i,0}$ plus the Faraday rotation angle of the i-th element (see Eq.~2).

The line-of-sight component of the magnetic field at the $i$-th element is $B_{i,\parallel}$, $P_i$ is the polarized intensity, $n_{e}$ is the thermal
electron density, and $\nu$ is the frequency of observation. The total Stokes $I$, Stokes $Q$, Stokes
$U$, and $\text{FD}$ are then found by summing the volume elements, $i$,
along the line of sight. The resulting output are Healpix images \citep{Gorski:2005ku} for each Stokes $I$, $Q$, and $U$ parameter. These synthetic observables can then be used to create a synthetic polarized fraction map.

\begin{table*}
\centering

\begin{tabular}{|c|c|c|c||c||c|c|c||c||c|c|c|c|c|}
\hline 
 & \multicolumn{3}{c||}{In fixed frame} & & \multicolumn{4}{c||}{For an observer looking North} & & \multicolumn{4}{c|}{For an observer looking South}\tabularnewline
\hline 
Model & $B_\phi$ & $B_z$ & $H_j$ &  & Model & $B_{\perp, \phi}$ & $B_{\parallel,z}$ & $C_{0}$ &  & Model & $B_{\perp, \phi}$ & $B_{\parallel,z}$ & $C_{0}$\tabularnewline
\hline 
\hline 
0 & -1.4 & 0 & 0 & & 0N & CCW & 0 & 0 & & 0S & CW & 0 & 0\tabularnewline
\hline 
1 & -1.4 & -0.5 & >0 & & 1N & CCW & >0 & <0 & & 1S &CW & <0 & >0\tabularnewline
\hline 
2 & 1.4 & 0.5 & >0 & & 2N & CW & <0 & >0 & & 2S & CCW & >0 & <0\tabularnewline
\hline 
3 & -1.4 & 0.5 & <0 & & 3N & CCW & <0 & >0 & & 3S & CW & >0 & <0\tabularnewline
\hline 
4 & 1.4 & -0.5 & <0 & & 4N & CW & >0 & <0 & & 4S & CCW & <0 & >0\tabularnewline
\hline 
\end{tabular}
\caption{Parameters for the five cases of the simple helix model. Models 1 and 2 have right-handed helicity ($H_j>0$) while Models 3 and 4 have left-handed helicity  ($H_j<0$). We use $B_{\parallel,z}$ to specify the line-of-sight component of $B_z$ alone (i.e., excluding any contribution from $B_\phi$). We use the usual convention that $B_{\parallel,z}>0$ if the magnetic field is directed towards the observer. This is distinct from the sign of $B_z$ since that is defined for a fixed $z$-axis, which does not depend on the observer's location. We use $B_{\perp,\phi}$ to mean the observer's plane-of-sky projection of $B_\phi$. The direction of $B_{\perp,\phi}$ depends of the observer's location, where the sign of $B_\phi$ does not. Here CW refers to a clockwise direction, and CCW as counter-clockwise. The sign of  $C_0$ applies at high frequencies, where Faraday rotation is negligible and for situations where the value of $B_{\text{rms}}$ is moderate.}
\label{tab:helix-cases}
\end{table*}

\begin{figure*}
\centering 

\vspace{0.5cm}

\includegraphics[width=17cm]{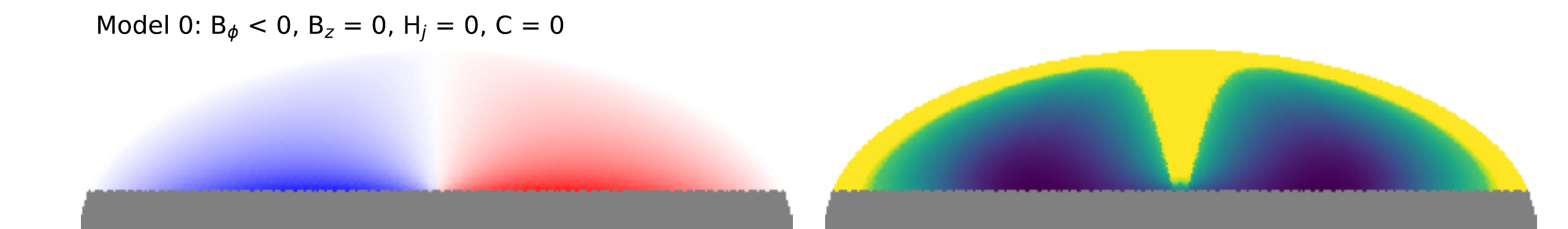}

\vspace{0.5cm}

\includegraphics[width=17cm]{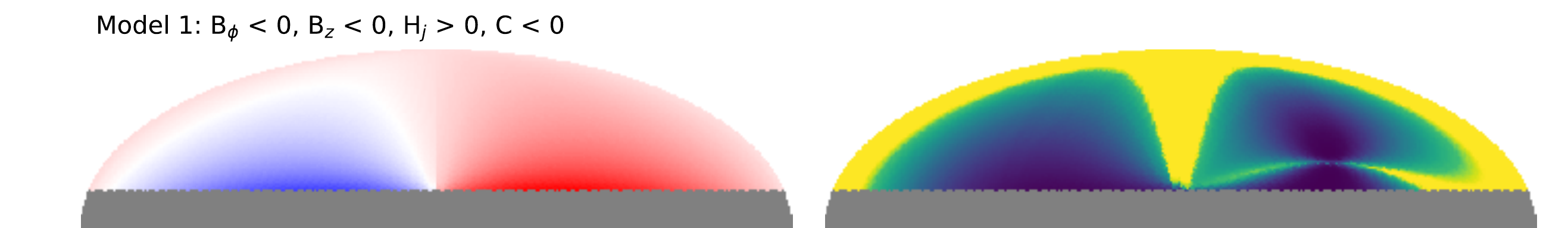}

\vspace{0.5cm}

\includegraphics[width=17cm]{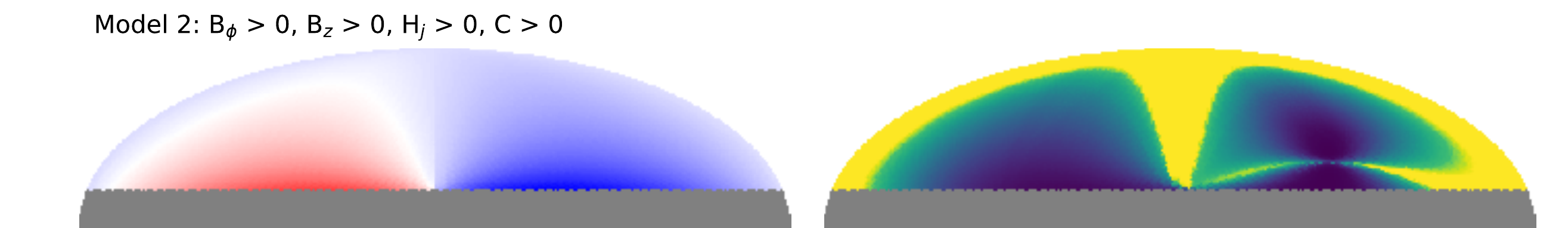}

\vspace{0.5cm}

\includegraphics[width=17cm]{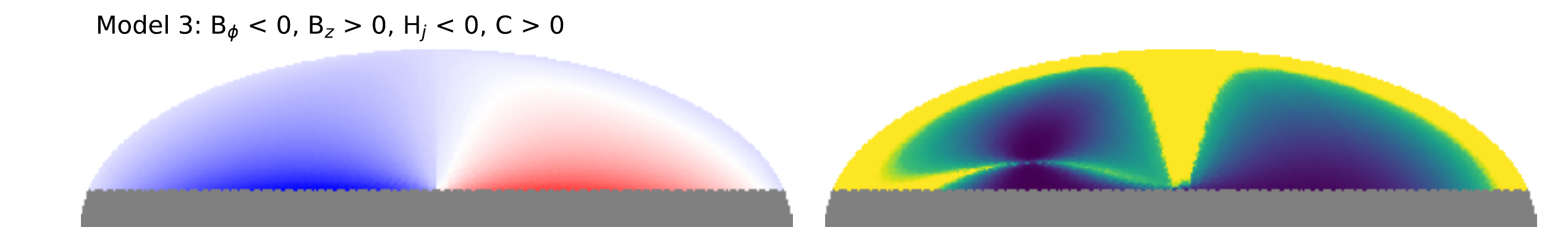}

\vspace{0.5cm}

\includegraphics[width=17cm]{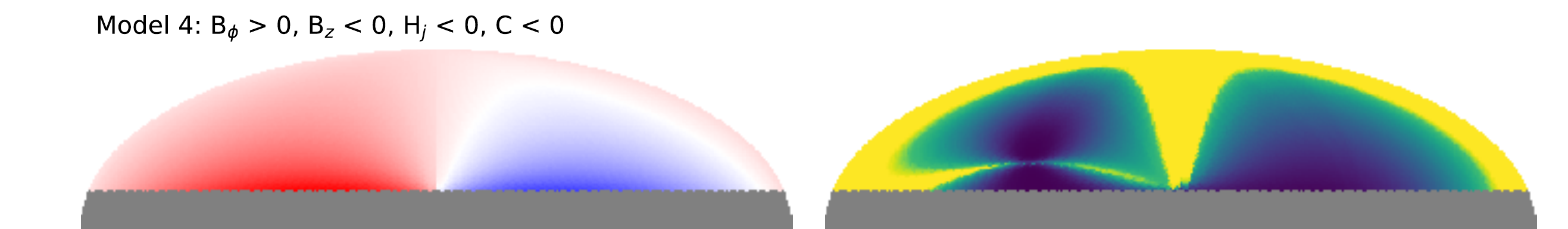}

\vspace{0.5cm}

\includegraphics[width=17cm]{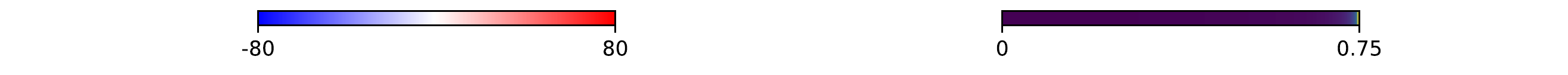}
\begin{scriptsize}\caption{\label{fig:simple-helix}FD [rad~m$^{-2}$] (left) and polarized fraction (right) for the 5 cases of a simple helical field described in Table~\ref{tab:helix-cases}. These are simulated observations of the Northern hemisphere (Galactic latitude, $b>0$) of a model galaxy, with an observer located at the Sun's position. They are shown in Galactic coordinates in Mollweide projection, with Galactic longitude $l=0$ at the centre. In all cases, the region for $|b|<15^\circ$ has been masked out as it is not included in these analyses. Polarized fraction is shown for the high-frequency limit ($\nu=30$~GHz) where Faraday rotation is negligible. The $\text{FD}$ colour scale is saturated at $|\text{FD}|=80$~rad~m$^{-2}$. More than 90\% of $\text{FD}<80$~rad~m$^{-2}$. The maximum  $\text{FD}=158$~rad~m$^{-2}$. Note that the colour scale for the polarized fraction appears very skewed towards the theoretical limit ($\simeq0.75$) since it has been histogram equalized. This is because the models shown have $B_{\text{rms}}=0$~$\mu$G, and most of the polarized fraction values are very close to this limit. However there are still some values very near 0, which can only be easily seen with this extreme colour map.
}\end{scriptsize} 

\end{figure*}

\begin{figure*}
\centering 
\begin{minipage}{8.5cm}
\includegraphics[width=8.4cm]{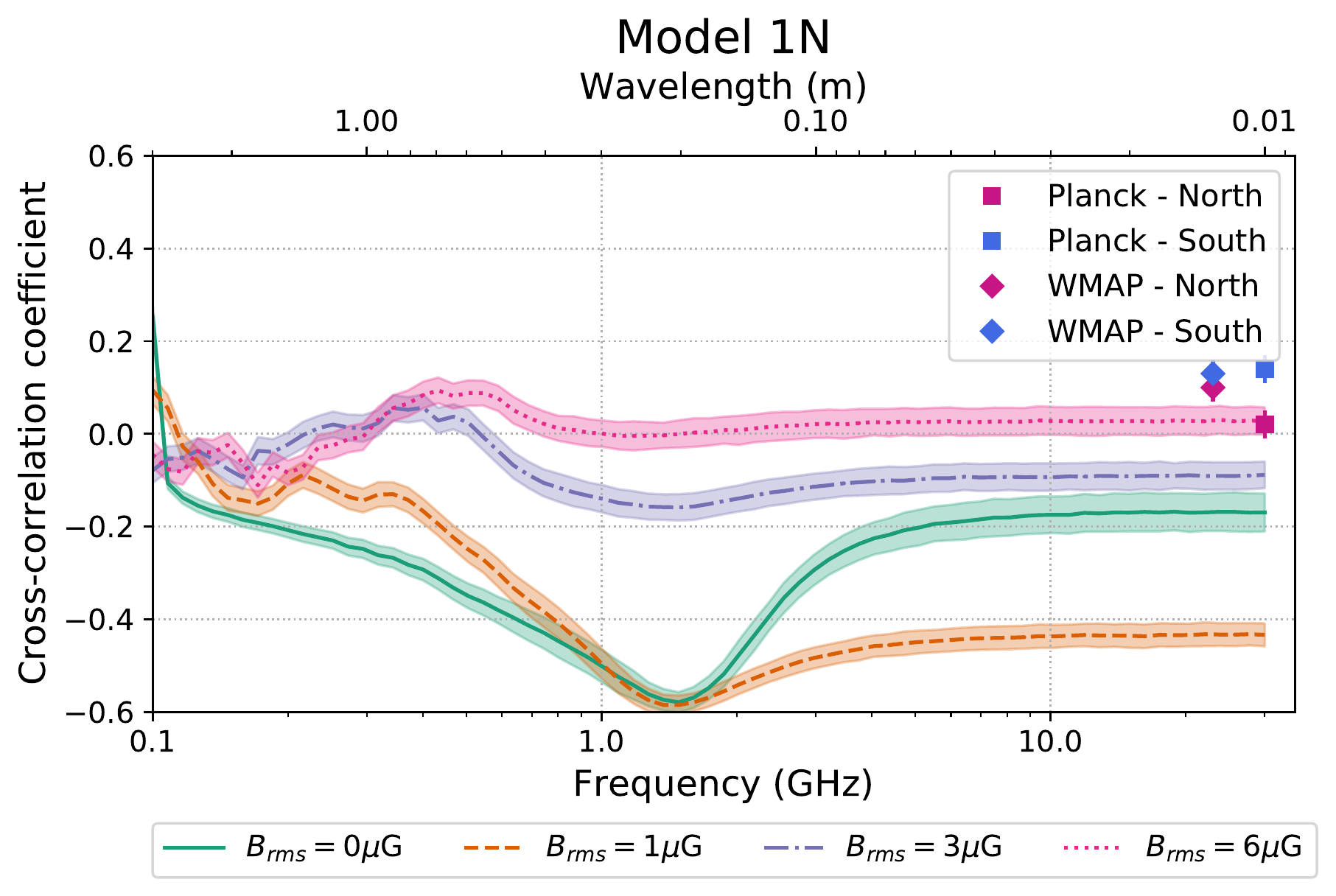}
\end{minipage}
\hfill
\begin{minipage}{8.5cm}
\includegraphics[width=8.4cm]{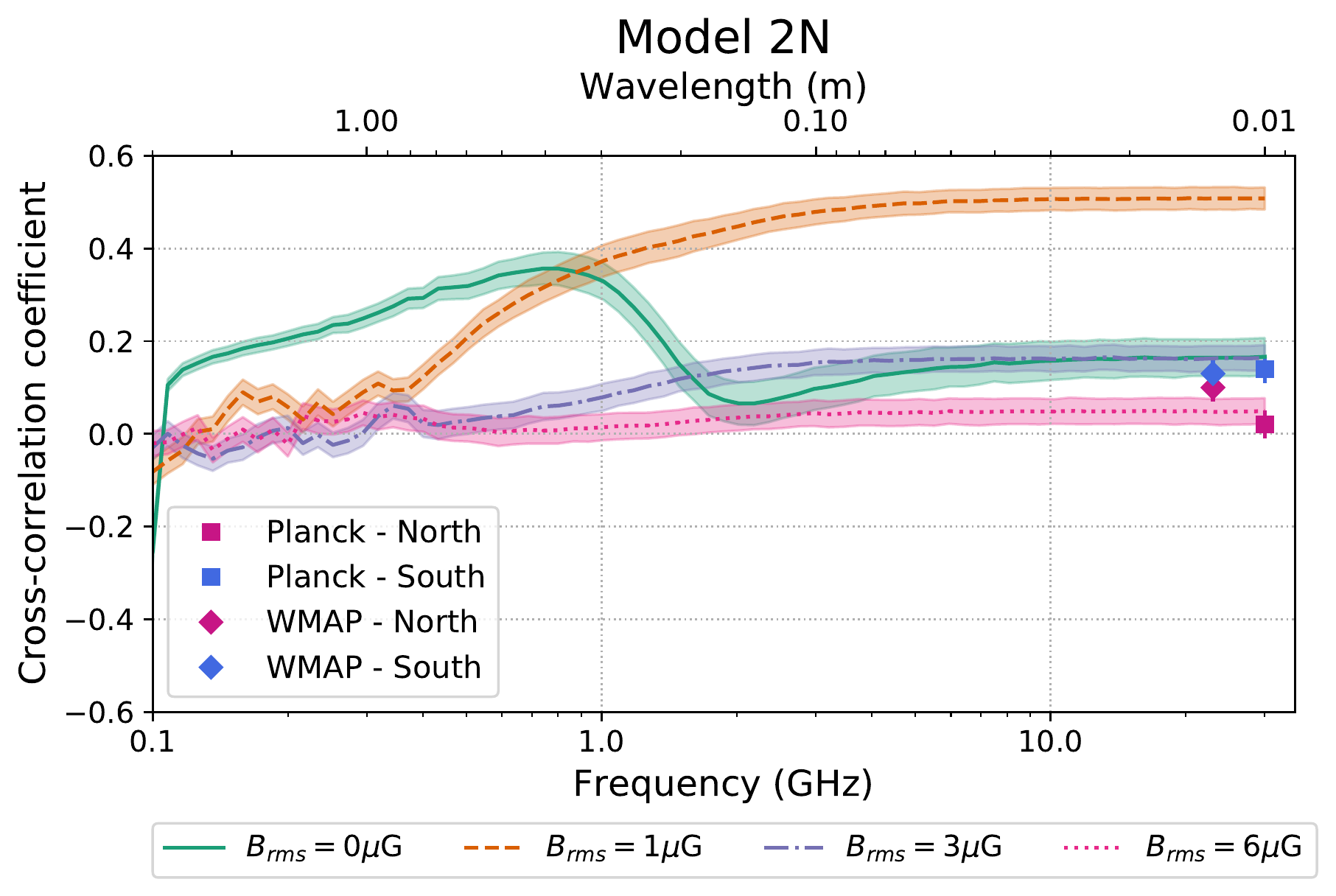}
\end{minipage}
\begin{minipage}{8.5cm}
\includegraphics[width=8.4cm]{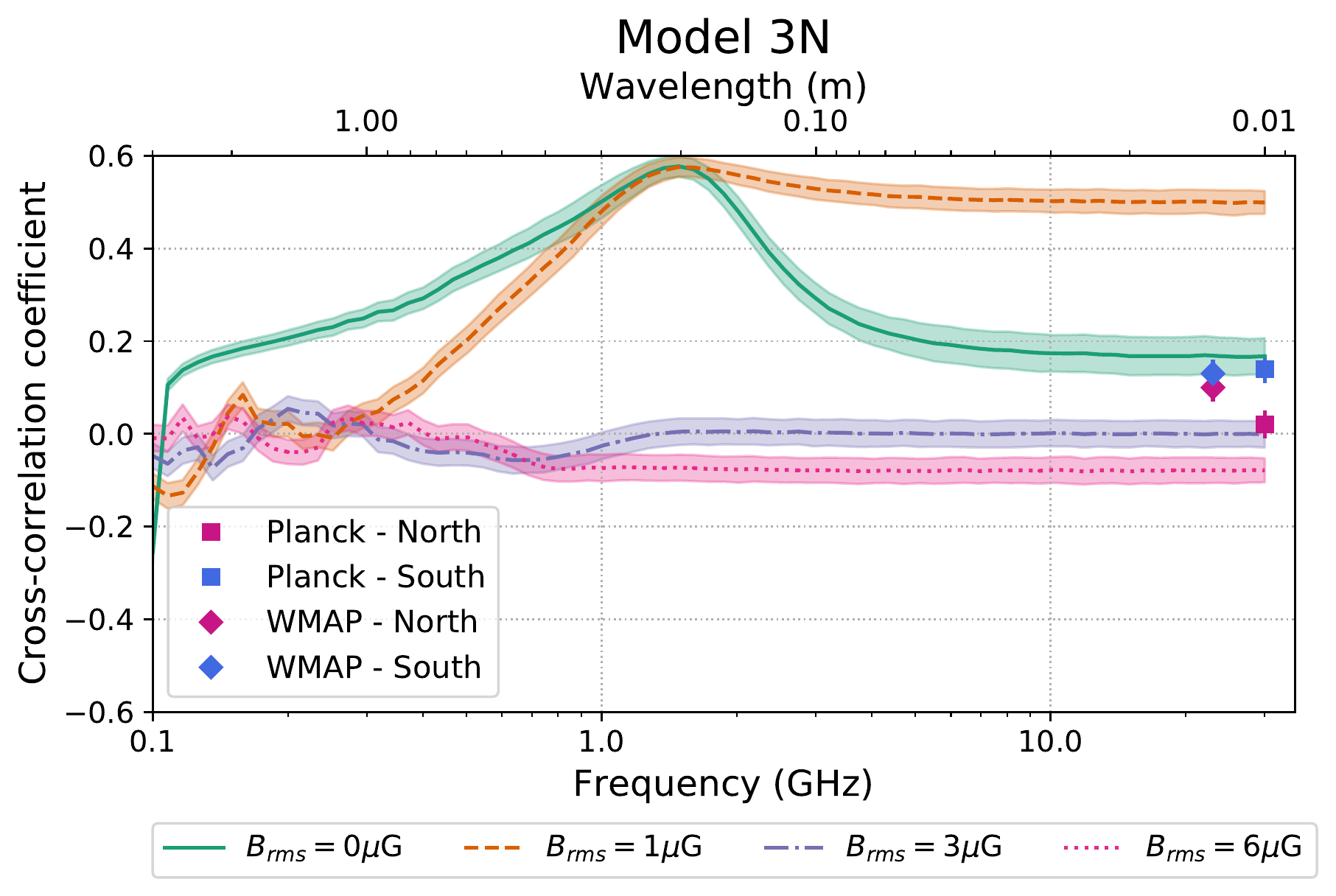}
\end{minipage}
\hfill
\begin{minipage}{8.5cm}
\includegraphics[width=8.4cm]{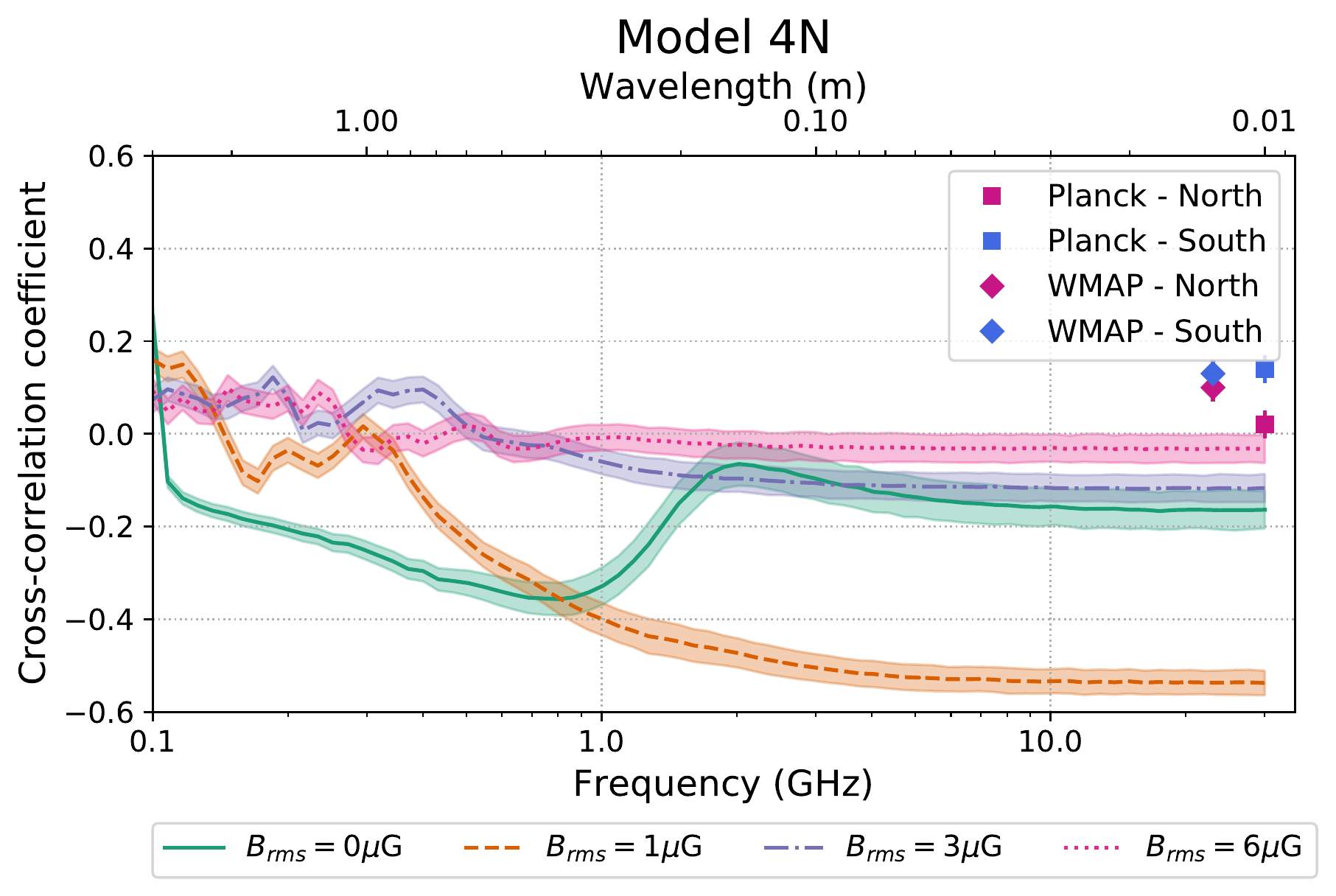}
\end{minipage}
\begin{scriptsize}\caption{\label{fig:freq-dependence}Frequency dependence of the cross-correlation coefficient, $C$, in the four cases of a simple helical field as described in Table~\ref{tab:helix-cases}, and as observed for the Northern hemisphere, plus the data points from high-frequency observations calculated in Sec~\ref{sec:data}. Models 1 and 2 (top row) both have right-handed helicity ($H_j>0$), while Models 3 and 4 (bottom row) have left-handed helicity ($H_j<0$). The models are shown with and without the random magnetic field components as described in Sec.~\ref{sec:toymodels}. The 1$\sigma$ uncertainty, shown by the shaded region, is derived from the bootstrap method used for the data and described in Sec.~\ref{sec:data}. } 
\end{scriptsize} 

\end{figure*}

For each model we use the total FD for the model and polarized fraction, $P/I$, to calculate $C$. In Sec.~\ref{sec:toymodels} we use a simple helical field model and compute $C$ for the hemisphere of the model where $z>0$ (i.e., the Northern hemisphere in the convention of the data). We do this for clarity in dealing with the cases individually, but we explain how these results apply to the Southern hemisphere. In Sec.~\ref{sec:dynamo}, we compute $C$ for both hemispheres using a somewhat more realistic, physically motivated model. In both models, we exclude the plane ($|b|\leqslant15^\circ$) for the computation of $C$, which is the same as how we treat the data (see Sec.~\ref{sec:data}). We also use the same bootstrapping method as described in  Sec.~\ref{sec:data} to calculate the mean and $1\sigma$ uncertainties for $C$.

\subsection{Simple Helical Field}
\label{sec:toymodels}
We first investigate whether a simple, toy model of a large-scale helical field can reproduce an analagous correlation between FD and polarized fraction as seen in the result of \citet{Volegova:2010go}. We construct a helical field with
\begin{equation}
{\bf{B}}=B_{\phi}{\widehat {\bf{e}}_\phi}+B_{z}{\widehat {\bf{e}}_z },
\end{equation}
where $\phi$ is the azimuthal angle around the $xy$-plane , $B_{\phi}$ is the azimuthal magnetic field component, $B_{z}$ is the magnetic field component perpendicular to the plane of the Galaxy, and ${\widehat {\bf{e}}_\phi}$ and ${\widehat {\bf{e}}_z}$ are the corresponding unit vectors. 

For simplicity, we use a constant thermal electron density, $n_e=0.01$~cm$^{-3}$, which is consistent with the average value in the Galactic disk \citep{2017ApJ...835...29Y}. The CRE model defines the spectral index and the CRE spatial density distribution at all points in the volume. We use $p=-3$ (in Eq.~\ref{eqn:hammurabi}) as this is the typical value used in other Galactic models \citep{Collaboration:2016eh}. For the spatial distribution of CREs, we use a simple exponential disk with a scale height, $h_d=1$~kpc and a radial scale length, $h_r=5$~kpc, as was used for the WMAP model \citep{Page:2007ce}. We find that the value of $C$ is very sensitive to the CRE density at high-latitudes, but not very sensitive to the thermal electron density. This is why we apply a scale height to the CRE model, but use a constant value for the thermal electron density.

The magnitude of the coherent magnetic field is chosen to be $\bf{|B|}$$=1.5~\mu$G, with $|B_{\phi}|=1.4~\mu$G and $|B_{z}|=0.5~\mu$G. The current helicity for this model is found using Eq.~\ref{eqn:helicity}, and is given by 
\begin{equation}
H_j = (B_\phi B_z)/r.
\end{equation}
Thus, when $B_\phi$ and $B_z$ have the same sign, $H_j>0$ (right-handed helicity) and when $B_\phi$ and $B_z$ have opposite sign, $H_j<0$ (left-handed helicity).

We model five cases described in Table~\ref{tab:helix-cases} for a range of frequencies, $0.1<\nu<30$~GHz, from the low-frequency case where the Faraday rotation is large ($\Delta\chi\simeq515^\circ$, see Sec.~\ref{sec:intro}), to the high-frequency case where Faraday rotation is negligible ($\Delta\chi<1^\circ$).

Since the observer in this model is positioned at $z=0$ (i.e., in the Galactic plane), and since we only consider the Northern hemisphere in this section, the sign of $B_z$ tells us whether the $z$-component of the field is pointed towards or away from the observer. 

This can be understood by considering that in the instances of this simple helical field, we define the same field throughout the entire box, with the observer located at the centre plane of the box (i.e., at $z=0$). Models 1 and 2 are both right handed fields. The difference between them is simply the viewing angle. For Model 1 you can imagine looking at your right hand with your thumb pointing down (defining $B_z$) where your fingers appear to curl clockwise (CW, defining $B_{\phi}$). In Model 2 you can imagine viewing your right hand with your thumb pointing up where your fingers appear to curl counter-clockwise (CCW). The difference for the two hemispheres is analogous: in the Northern hemisphere we view the field from below, which we describe as Model 2N in Table~\ref{tab:helix-cases} (i.e., like holding your right hand above your head while keeping your thumb pointing down), while in the Southern hemisphere, Model 2S, we view the same field but from above.

For Model 1 we have $B_z<0$ and $B_\phi<0$, which for the North, Model 1N, has $B_{\text{sky}, \phi}=$~CCW and $B_{\text{los},z}>0$ (i.e., pointed towards the observer). For this same field in the Southern hemisphere, Model 1S the observer sees the opposite, $B_{\text{sky}, \phi}=$~CW and $B_{\text{los},z}<0$ (i.e., pointed away from the observer), which is like Model 2N. The total field has $H_j>0$ and this remains constant in the two hemispheres. 

The first of these cases, Model 0, has $B_z=0$, and thus represents a purely toroidal field. In this case, there is no helicity ($H_j=0$) and we find that the Eastern (left) side of the model observation has a negative $\text{FD}$ (field is directed away from the observer) and the Western (right) side of the model observations has a positive $\text{FD}$ (field is directed towards the observer), as shown in the top row of Fig~\ref{fig:simple-helix}. In this case, we find $C=0$ for all frequencies.

The next four cases, Models 1-4, have $B_z \not= 0$, so that the field becomes a helical corkscrew. In these cases we find the $\text{FD}$ distribution becomes asymmetric due to the additional $z$-component and the Sun's off-centre position in the Galaxy. Moreover, the polarized fraction is also asymmetric, even in cases of high-frequency observations where the Faraday rotation is very small, as shown in Fig.~\ref{fig:simple-helix}. This asymmetry is due only to geometric cancellations of the magnetic field along the line of sight due to the twist introduced by the helicity, and for most frequencies, $C\ne0$, as shown in Fig.~\ref{fig:freq-dependence}. In Table~\ref{tab:helix-cases}, we give the sign of $C$ at high frequencies, where Faraday rotation is negligible (i.e., for $\lambda=0$), which we call $C_0$. Other values of $|B_{z}|$ were explored and the trend remains the same in all cases, however the values of $C$ differ. We choose to focus on a single value here, but this could be further explored in future work.

This correlation can be understood by noting that e.g., in Model 1 (2nd row in Fig.~\ref{fig:simple-helix}), high polarized fraction (yellow) regions tend to have small positive $\text{FD}$ (light red), while large positive $\text{FD}$ (dark red) regions tend to have low polarized fraction (dark blue), and hence a negative contribution to the correlation.

If we flip the handedness of the helicity by flipping the sign of $B_\phi$ while keeping the sign of $B_z$ unchanged (i.e., switching between Models 1 and 4 or between Models 2 and 3), then the east-west pattern of the $\text{FD}$ distribution flips (i.e., reflected about longitude $l=0^\circ$), but so does the pattern of the polarized fraction distribution. In this case, the sign of $C_0$ does not change.

However, if we flip the handedness of the helicity by flipping the sign of $B_z$ while keeping the sign of $B_\phi$ unchanged (i.e., switching between Models 1 and 3 or between Models 2 and 4), then the east-west patterns of the $|\text{FD}|$ and polarized fraction distributions flip (as in the previous case), but in contrast to the previous case, the general sign of $\text{FD}$ itself (e.g., negative east and positive west) doesn't flip.

We also investigate how these cases change when we introduce a normalized Gaussian random component, $B_{\text{rms}}$. We test this using values of $B_{\text{rms}}=1$, $3$, and $6$~$\mu$G. A value of $B_{\text{rms}}=6$~$\mu$G is approximately what we expect for the magnitude of the random component in the Milky Way Galaxy \citep{Haverkorn:2006bb}. Nominally, this random component should have a Kolmogorov power spectrum with maximum scales $\sim100$~pc. However, due to the pixel-scale of our model, the maximum scale is set at $200$~pc (so that it is sampled by $\sim3$~pixels). It is not feasible to include other turbulent scales at this model resolution, so this is effectively single-scale turbulence. The magnitude of the random component is modulated by a simple exponential disk with a scale height, $h_d=1$~kpc and a radial scale length, $h_r=10$~kpc, as was implemented for other Galactic magnetic field models in \citet{Collaboration:2016eh}.

\begin{figure*}
\centering 
\begin{minipage}{16.5cm}
\includegraphics[width=16.4cm]{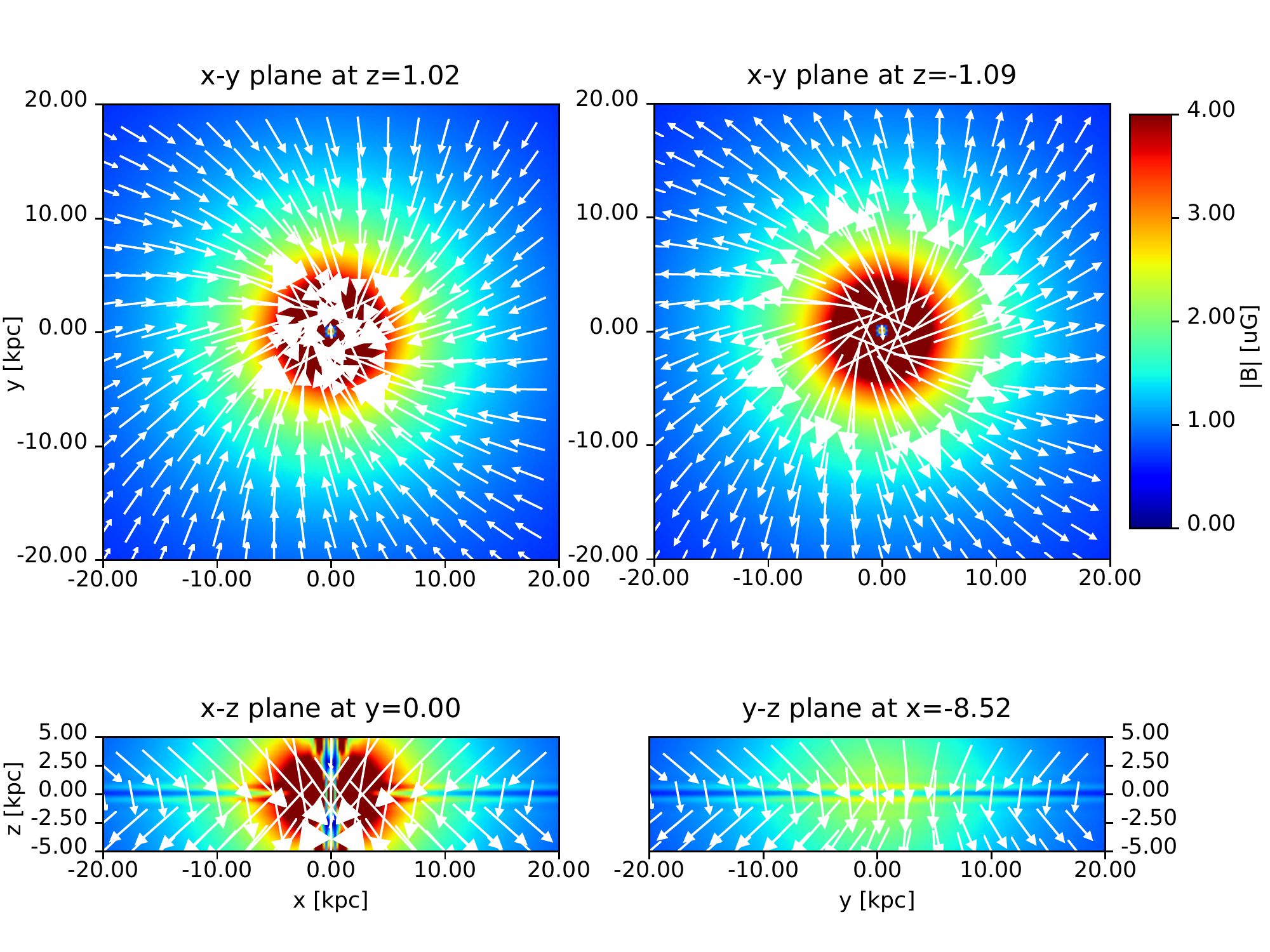}
\end{minipage}

\begin{scriptsize}\caption{\label{fig:fields} Plot of the dipolar case of the magnetic field from the dynamo model ($m=0$ mode) described in Sec.~\ref{sec:dynamo}.  Top left: $x-y$ plane (top-down view) cut through $z=1.02$~kpc (Northern hemisphere). Top right: $x-y$ plane (top-down view) cut through $z=-1.09$~kpc (Southern hemisphere). Bottom left: $x-z$ plane (side view) cut through $y=0$ (i.e., through the galactic centre). Bottom right: $y-z$ plane (side view) cut through $x=-8.52$~kpc (i.e., through the position of the observer).  The colourbar and the length of the arrows in the plot are scaled according to the total magnitude of the magnetic field.}
\end{scriptsize} 

\end{figure*}

\begin{figure*}
\centering 
\begin{minipage}{16.5cm}
\includegraphics[width=16.4cm]{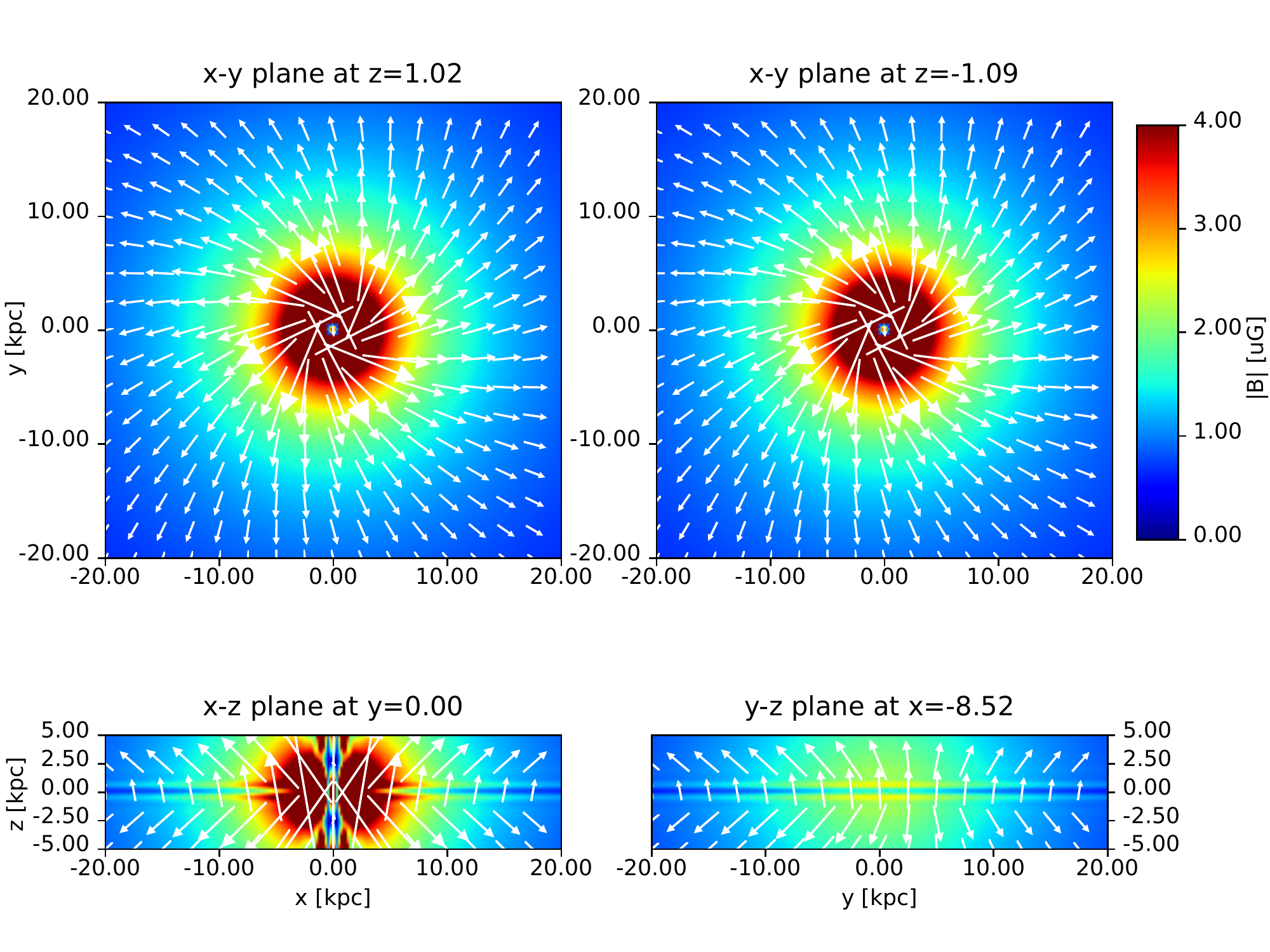}
\end{minipage}

\begin{scriptsize}\caption{\label{fig:quad} Plot of the quadrupolar case of the magnetic field from the dynamo model ($m=0$ mode) described in Sec.~\ref{sec:dynamo}.  Top left: $x-y$ plane (top-down view) cut through $z=1.02$~kpc (Northern hemisphere). Top right: $x-y$ plane (top-down view) cut through $z=-1.09$~kpc (Southern hemisphere). Bottom left: $x-z$ plane (side view) cut through $y=0$ (i.e., through the galactic centre). Bottom right: $y-z$ plane (side view) cut through $x=-8.52$~kpc (i.e., through the position of the observer).   The colourbar and the length of the arrows in the plot are scaled according to the total magnitude of the magnetic field. 
}
\end{scriptsize} 

\end{figure*}

The results, summarized in Fig.~\ref{fig:freq-dependence}, apply to an observer looking towards the Northern hemisphere. These can also apply to the Southern hemisphere, as discussed previously in Sec.~\ref{sec:modelling}. 

We note the following:
\begin{enumerate}
    \item Models 1 and 3 have the same $B_\phi$ but opposite $B_z$, and therefore opposite $H_j$. This is why, in the case $B_{\text{rms}} = 0~\mu$G (green, solid curve), they have opposite $C_{0}$ (because of opposite $B_z$). More generally, for each value of $B_{\text{rms}}$, they have opposite $|C(\nu)|$. Likewise for Models 2 and 4.
    \item When $B_{\text{rms}}=0~\mu$G (green, solid curve), $C$ in each model has a single sign for all frequencies, except for very low frequencies where there is a great deal of Faraday rotation. This sign is consistent with the sign of $C_{0}$ indicated in Fig.~\ref{fig:simple-helix}.
    \item When $B_{\text{rms}} = 0~\mu$G or $1~\mu$G, Models 1 and 3 have a distinct peak in $|C|$ at $\nu\approx1.5$~GHz, which is close to the peak frequency of $\nu=2$~GHz found by \citet{Volegova:2010go}.
    \item When $B_{\text{rms}}=1~\mu$G, the value of $|C_{0}|$ is larger than for $B_{\text{rms}}=0~\mu$G. In this case, the coherent component of the magnetic field, which has a magnitude of $1.5~\mu$G, is larger than the random component. Therefore, the total magnetic field does not change sign. The larger value of $|C_{0}|$ can be explained by considering that the amount of depolarization increases as the random component increases. Since the total magnetic field does not change sign, the $\text{FD}$ also does not change sign. Thus, there is a stronger correlation between $\text{FD}$ and polarized fraction when compared to the $B_{\text{rms}}=0~\mu$G case.
    \item When $B_{\text{rms}}$ is larger than the coherent component (i.e., $B_{\text{rms}}=3~\mu$G and $B_{\text{rms}}=6~\mu$G), the value of $|C|$ tends to be smaller than for $B_{\text{rms}}=0~\mu$G or $1~\mu$G (at most frequencies). Here the random component dominates the coherent component. Since the total magnetic field can randomly change directions, the $\text{FD}$ randomly changes sign as a function of position, so there is a weaker correlation. 
      
\end{enumerate}

\begin{figure*}
\centering 
\begin{minipage}{17cm}
\includegraphics[width=17cm]{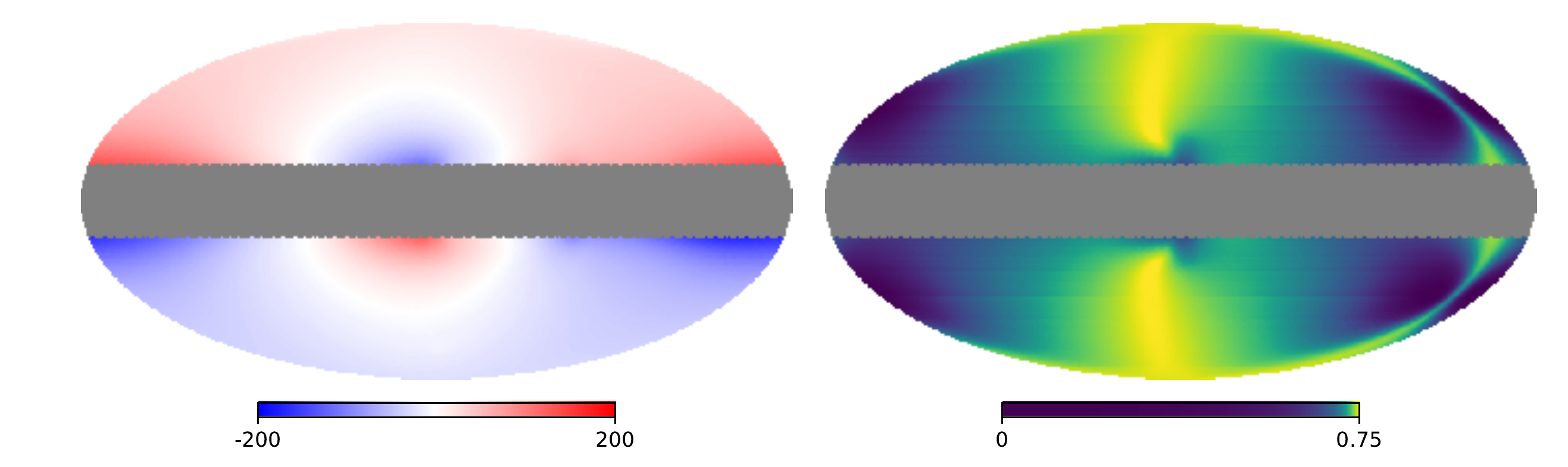}
\end{minipage}
\hfill
\begin{minipage}{17cm}
\includegraphics[width=17cm]{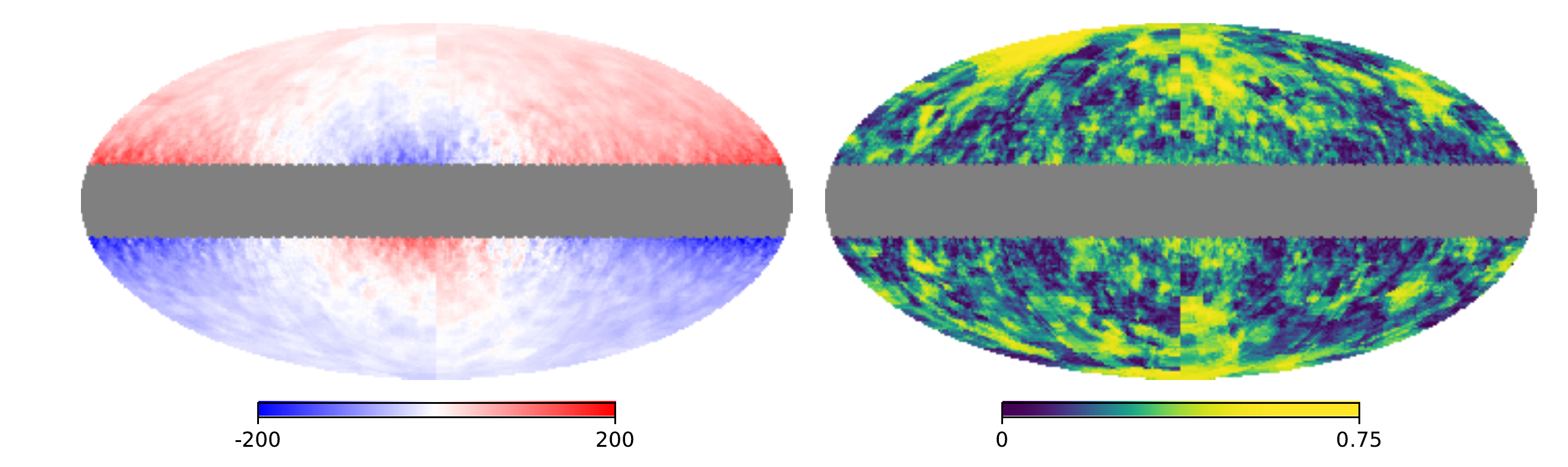}
\end{minipage}

\begin{scriptsize}\caption{\label{fig:modelmapsm0-dipole}Top row: Simulated $\text{FD}$ [rad~m$^{-2}$] (left) and polarized fraction (right) at $\nu=30$~GHz for the dipolar case of the dynamo model shown in Fig.~\ref{fig:fields} and a magnetic field with $B_{\text{rms}}=0$~$\mu$G. The orientation of all maps is the same as the observations in Fig.~\ref{fig:wmap}, with the Galactic centre at the centre of the map and the Galactic plane oriented horizontally through the centre of each map. In all cases, the region for $|b|<15^\circ$ has been masked out as it is not included in these analyses. Bottom row: Same as for the top row, but for $B_{\text{rms}}=6$~$\mu$G.}
\end{scriptsize} 

\end{figure*}

We have shown that a simple, toy model of a large-scale helical field does produce correlations between FD and polarized fraction as seen in the result of \citet{Volegova:2010go}. However, they found that when $H_m>0$,  $C>0$ and we find the relationship here is more complex. In our simple, large-scale models, the sign of $C$ does not appear to be explicitly linked to the sign of $H_j$. Rather in these cases, $C$ has the opposite sign to $B_{\text{los},z}$. This is discussed further in Sec.~\ref{sec:modeldiscussion}.

\subsection{Dynamo model}
\label{sec:dynamo}
In order to test whether these correlations exist in a more physically motivated, and more complex magnetic field model, we use a spirally symmetric dynamo model
that was developed by  \citet{Henriksen:2017gp, Henriksen:2018uk} and applied to modelling NGC 4631 by \citet{2019MNRAS.487.1498W}. We test both dipolar and quadrupolar symmetries.

This model is a particular solution that was motivated by the desire to find cases in classical dynamo theory where there is agreement between the model and observed properties including magnetic disc spirals and X-shape poloidal fields. The model contains the alpha effect, a shearing outflow and diffusion in a `pattern' frame. The pattern frame is defined with respect to the spiral arms, which may have a different rotation rate than the mean disc rotation. This is really the magnetic spiral arm pattern speed, which may be different from the stellar arm speed. It is all in the context of scale invariance which has the merit of reproducing most of the known numerical effects with computational ease.

A particular model is parameterized using the variables  $m$, $a$, $u$, $v$, $w$, $T$, $q$, $\epsilon$, $C1$, and $C2$. The spiral mode is defined by $m$, where we use the axisymmetric $m=0$ mode, which has a diverging X-shaped morphology, which is similar to the halo magnetic field that is observed in external galaxies as discussed in Sec.~\ref{sec:intro}. The variable $a$ is a scaling parameter, and $u$, $v$, and $w$ are velocity components in the $x$,$y$, and $z$ directions, respectively. We use a case where $a= 1$, which conserves a global velocity. The $m=0$ mode that we explore here has no radial velocity nor circular velocity in the pattern frame ($u=v=0$), but an outflow ($w>0$). An outflowing vertical velocity component (sometimes called a fountain flow) is often included in galactic dynamo models \citep[e.g.,][]{2006A&A...448L..33S, 2014MNRAS.443.1867C}. Here we use $w=1$. The pitch angle of the spiral is set by the variable $q$. We use $q=4.9$, which corresponds to a pitch angle of $-11.5^\circ$, the value used in most Milky Way magnetic field models \citep{Collaboration:2016eh}. $T$ is a time variable and $\epsilon$ sets the rotation rate. We use a case that is observed at an arbitrary time taken to be the current epoch ($T=1$ and $\epsilon=-1$). Finally, $C1$, and $C2$ are  boundary conditions ($C1=0$ and $C2=1$). The model is briefly defined further in Appendix ~\ref{sec:dynamoappendix}, and a full description may be found in other work \citep{Henriksen:2017gp, Henriksen:2018uk, 2019MNRAS.487.1498W}.

We find the solution for this equation for points where $z>0$, and then assume a dipolar symmetry across the disk of the galaxy (i.e., $z=0$) to calculate points where $z<0$. Under this condition, $B_z$ is continuous across $z=0$, but $B_r$ and $B_\phi$ change sign. This necessarily means that $H_j$ also changes sign across $z=0$. We show this case plotted in Fig.~\ref{fig:fields}. 

We also model the case with quadrupolar symmetry, plotted in Fig.~\ref{fig:quad}. Here $H_j$ also changes sign across the midplane, as in the dipolar case. However, unlike the dipolar case, here $B_z$ changes sign while $B_r$ and $B_\phi$ keep the same sign across the midplane.

\begin{figure*}
\centering 
\begin{minipage}{8.5cm}
\includegraphics[width=8.4cm]{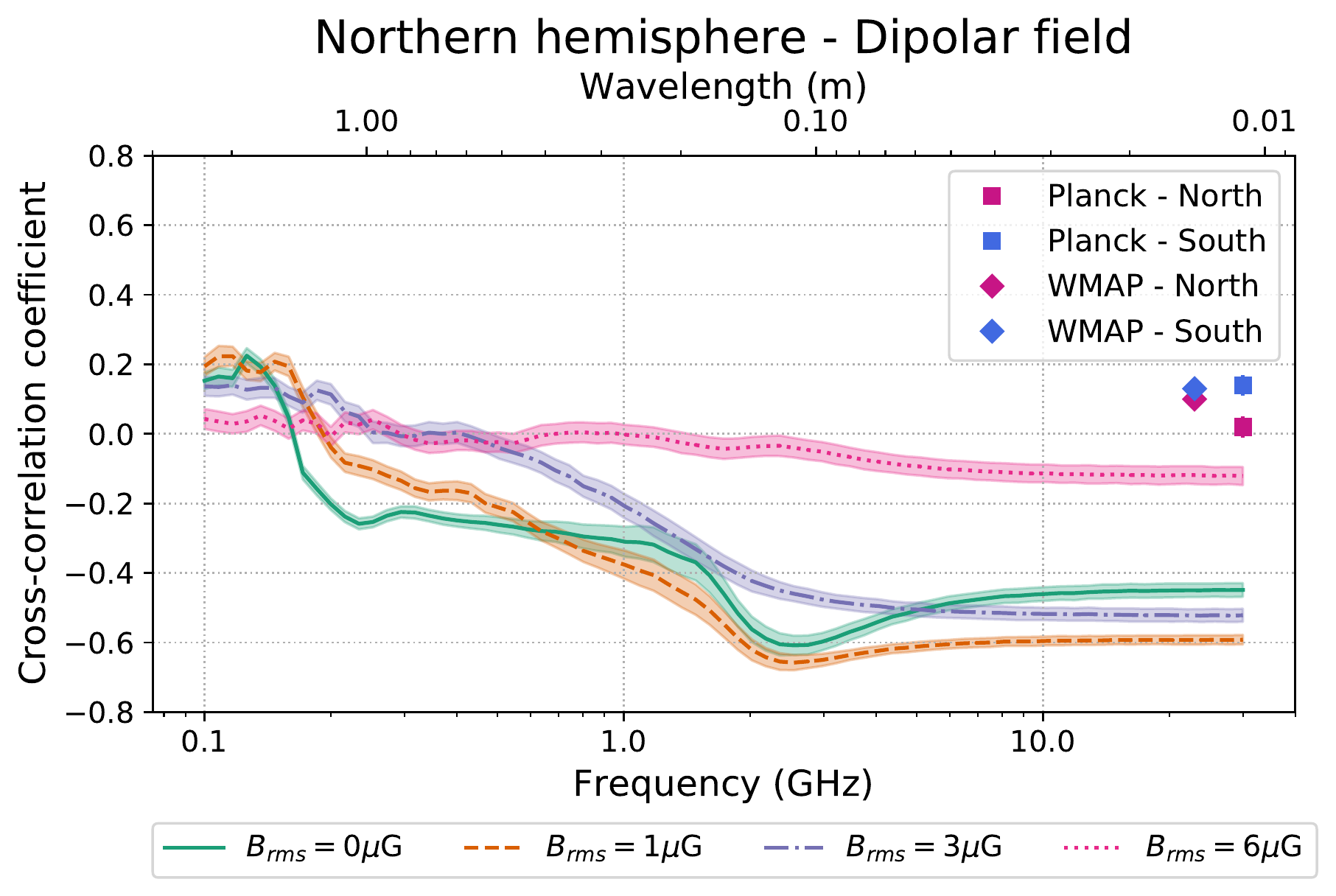}
\end{minipage}
\hfill
\begin{minipage}{8.5cm}
\includegraphics[width=8.4cm]{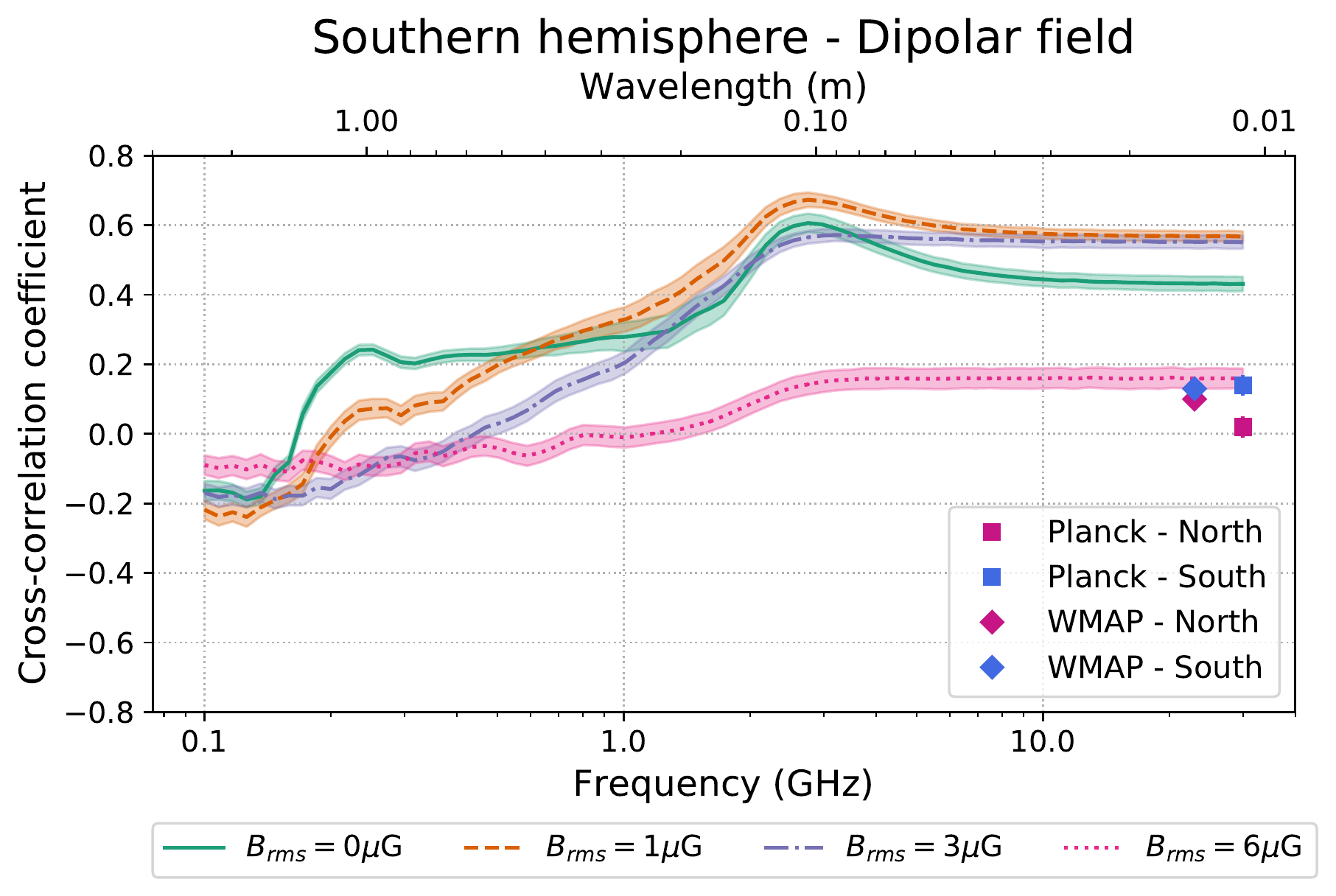}
\end{minipage}

\begin{scriptsize}\caption{\label{fig:freq-dependence-dynamo}Frequency dependence of the cross-correlation coefficient, $C$, for the dipolar case of the dynamo model, shown in Fig.~\ref{fig:fields}. This figure is presented in a similar way as for the simple helix model cases shown Fig.~\ref{fig:freq-dependence}. In this case, $B_{z}$ is pointing towards the observer in the Northern hemisphere (left), and $B_{z}$ is pointing away from the observer in the Southern hemisphere (right). 
}
\end{scriptsize} 

\end{figure*}

\begin{figure*}
\centering 
\begin{minipage}{17cm}
\includegraphics[width=17cm]{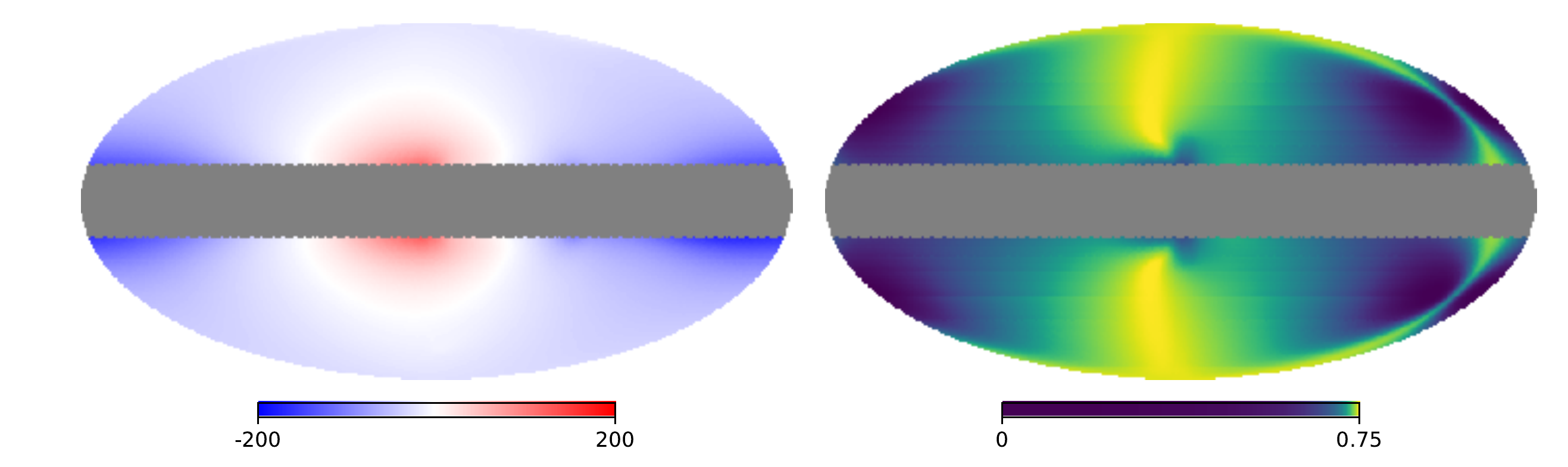}
\end{minipage}
\hfill
\begin{minipage}{17cm}
\includegraphics[width=17cm]{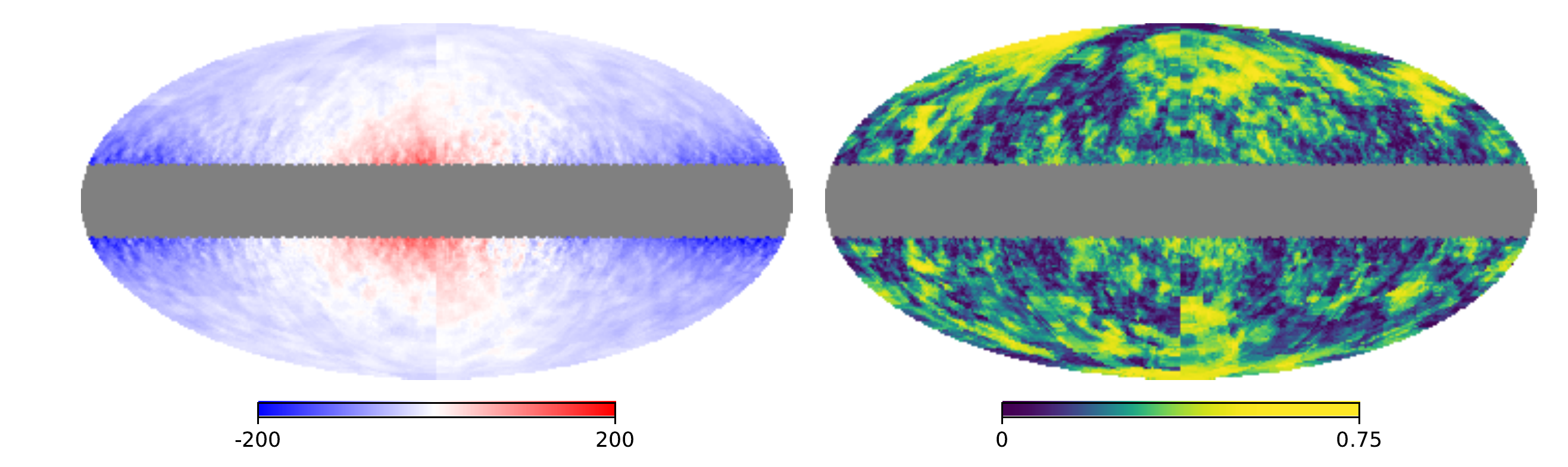}
\end{minipage}

\begin{scriptsize}\caption{\label{fig:modelmapsm0-quad}Top row: Simulated $\text{FD}$ [rad~m$^{-2}$] (left) and polarized fraction (right) at $\nu=30$~GHz for the quadrupolar case of the dynamo model shown in Fig.~\ref{fig:quad} and a magnetic field with $B_{\text{rms}}=0$~$\mu$G. The orientation of all maps is the same as the observations in Fig.~\ref{fig:wmap}, with the Galactic centre at the centre of the map and the Galactic plane oriented horizontally through the centre of each map. In all cases, the region for $|b|<15^\circ$ has been masked out as it is not included in these analyses. Bottom row: Same as for the top row, but for $B_{\text{rms}}=6$~$\mu$G.}
\end{scriptsize} 

\end{figure*}

\begin{figure*}
\centering 
\begin{minipage}{8.5cm}
\includegraphics[width=8.4cm]{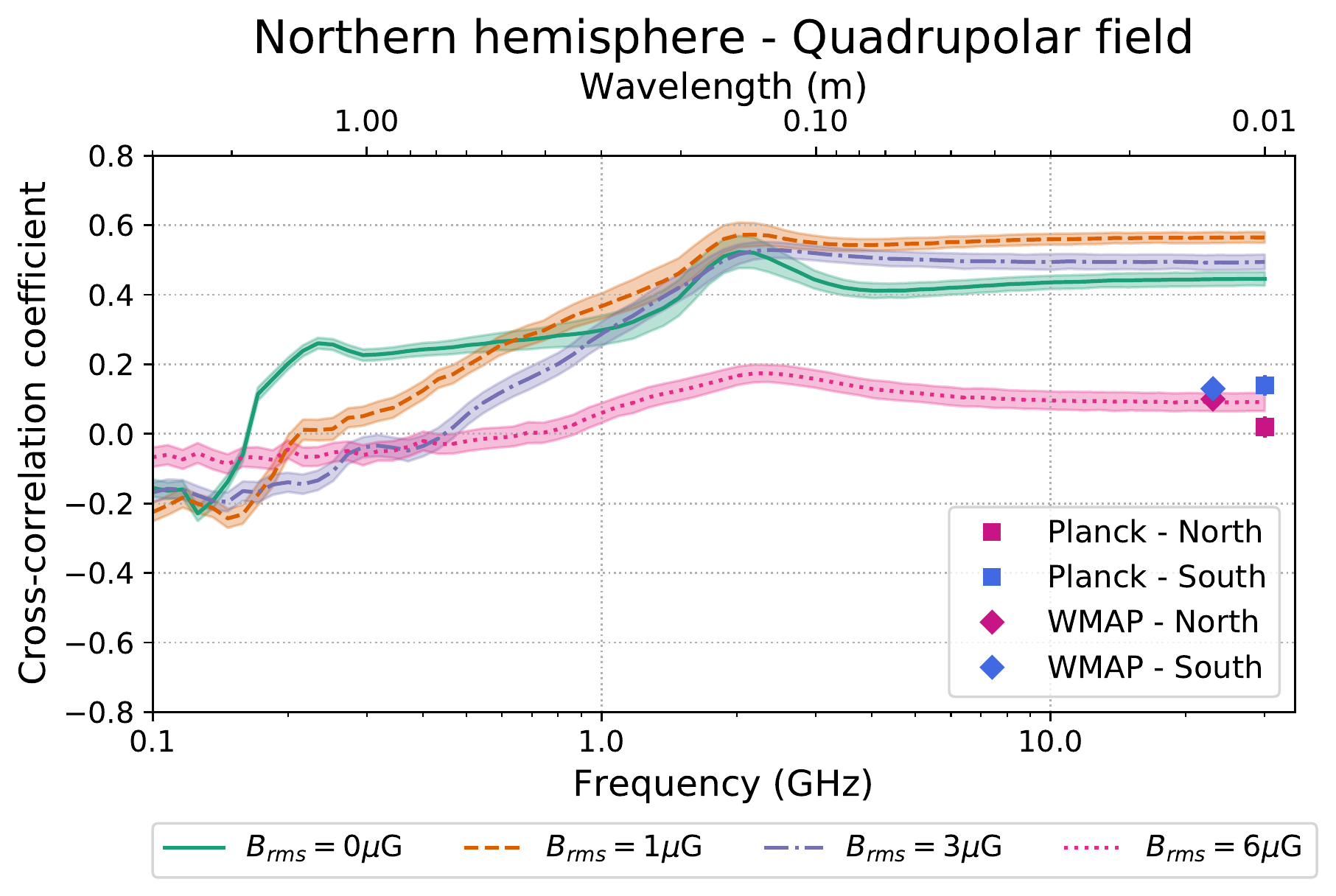}
\end{minipage}
\hfill
\begin{minipage}{8.5cm}
\includegraphics[width=8.4cm]{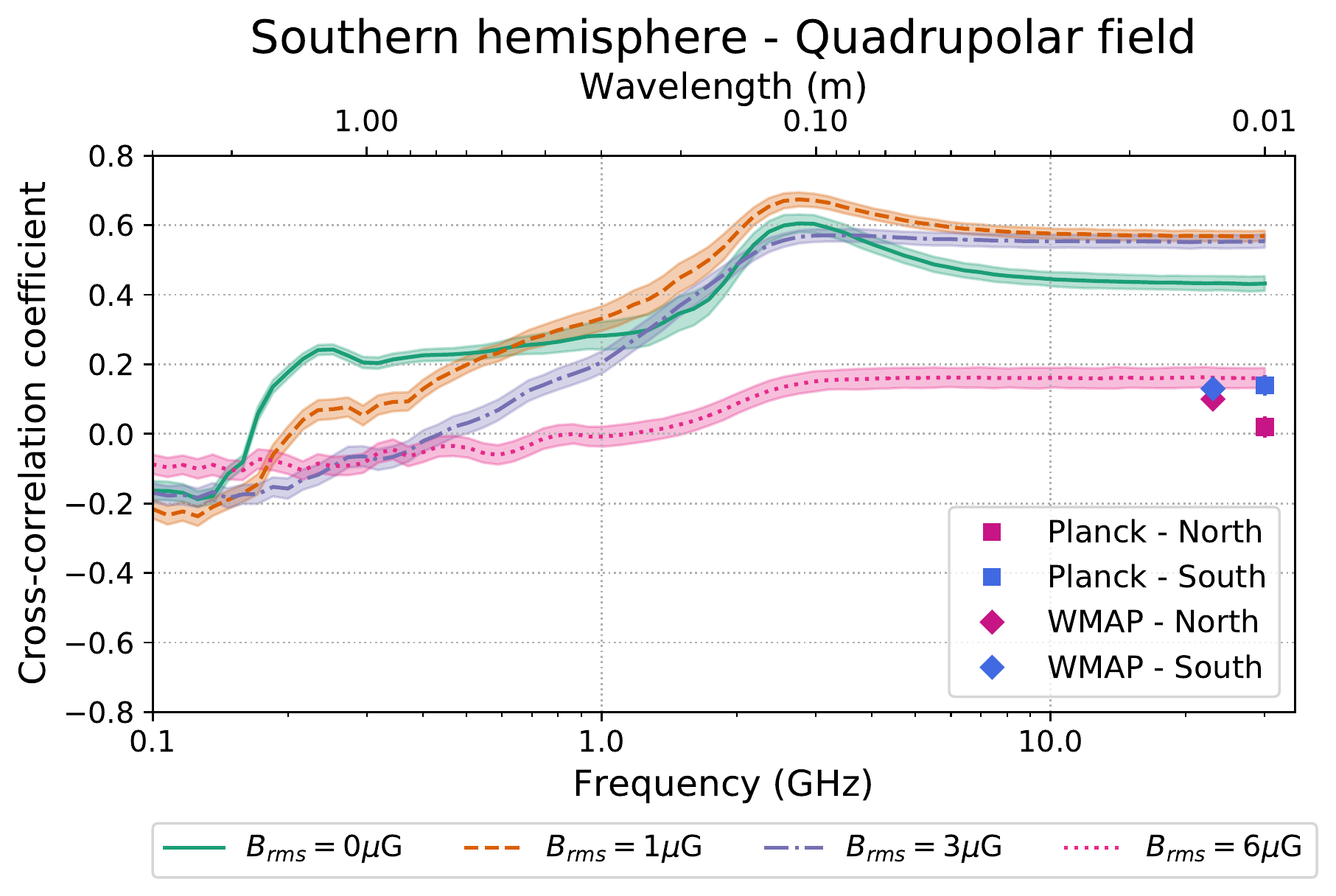}
\end{minipage}

\begin{scriptsize}\caption{\label{fig:freq-dependence-dynamo-quad}Frequency dependence of the cross-correlation coefficient, $C$, for the quadrupolar case of the dynamo model, shown in Fig.~\ref{fig:quad}. $B_{z}$ is pointing away from the observer in both hemispheres. }
\end{scriptsize} 

\end{figure*}

For convenience, we set to zero all values of the field immediately in the centre for $r<1$~kpc and within a vertical cone with opening angle, $\theta=10^\circ$. We do this because the field model diverges at the origin. Although not strictly physical, this only impacts a very small fraction of pixels ($<0.04\%$ of pixels) and this practice is consistent with what other field models have done where the field is very uncertain at the Galactic centre \citep[e.g., ][]{Jansson:2012ep}. We emphasize that we present this particular solution, for the dipolar and quadrupolar cases, as proof of concept and do not imply that these parameters represent the best fit for a model of the Milky Way's magnetic field.

To be consistent with our more physically motivated model, the simulated observations use the most recent Galactic thermal electron density model defined by \citet{2017ApJ...835...29Y}. We remove the variations in the electron density contribution that are added for local ionized features near to the Sun ($r<$ 1~kpc). We do this because at the resolution of our model, $\sim75$ pc per pixel, the local features are only a few pixels wide with sharp edges. As such, they produce significant artifacts in our simulated images. And since here we investigate the large-scale field, the local electron density is not relevant. However, the local electron density (and magnetic field) may be an important difference between these models and the observations and we discuss this further in Sec.~\ref{sec:discussion}. We use the same CRE distribution as described in the previous section.

As in the previous case, we compute the simulated observations for a range of frequencies, $0.1<\nu<30$~GHz, i.e., between large and negligible Faraday rotation regimes. Fig.~\ref{fig:modelmapsm0-dipole} shows the simulated observations for the high-frequency limit ($\nu=30$~GHz, $\lambda\sim0.01~$m) of the field model with dipolar symmetry (Fig.~\ref{fig:fields}), and Fig.~\ref{fig:freq-dependence-dynamo} shows the frequency dependence of $C$ for both the Northern and Southern hemispheres, still for the dipolar case. Similarly, Fig.~\ref{fig:modelmapsm0-quad} shows the simulated observations for the high-frequency limit of the field model with quadrupolar symmetry (Fig.~\ref{fig:quad}), and Fig.~\ref{fig:freq-dependence-dynamo-quad} shows $C(\nu)$ for the quadrupolar case.

At first sight, the $\text{FD}$ maps shown in Fig.~\ref{fig:simple-helix} on the one hand, and Figs.~\ref{fig:modelmapsm0-dipole} and  Fig.~\ref{fig:modelmapsm0-quad} on the other hand may appear quite different. This can be explained by considering that in the simple helix model, the field has no radial component ($B_r = 0$), so the $\text{FD}$ maps in Fig.~\ref{fig:simple-helix} are roughly symmetric with respect to longitude $l=0$ (centre of the maps), especially at low latitudes, where $B_z$ contributes little to $\text{FD}$. In contrast, in the dynamo model, the field has a significant radial component, so the $\text{FD}$ maps in Figs.~\ref{fig:modelmapsm0-dipole} and \ref{fig:modelmapsm0-quad} are roughly symmetric with respect to a non-zero longitude, whose value depends on the pitch angle.

For the dipolar model, the Northern hemisphere has $B_{\phi} < 0$ and $B_z < 0$, which suggests $H_j > 0$, and which has the same signs as Model 1. This is why the Northern hemisphere (left) plot in Fig.~\ref{fig:freq-dependence-dynamo}, is similar to the Model 1N plot in Fig.~\ref{fig:freq-dependence}. An important difference is that, as can be seen in Figs.~\ref{fig:fields} and \ref{fig:quad} (see the scale shown  in the colourbar), the coherent part of the magnetic field in the dynamo model is larger than in the simple helical model, which has a constant value of $1.5~\mu$G. Thus in Fig.~\ref{fig:freq-dependence-dynamo} we see the case where $B_{\text{rms}}=3~\mu$G is more similar to the case where $B_{\text{rms}}=1~\mu$G as compared to Fig.~\ref{fig:freq-dependence} where the $B_{\text{rms}}=3~\mu$G and $B_{\text{rms}}=1~\mu$G cases are quite different.

On the other hand, the Northern hemisphere of the quadrupolar field model has $B_{\phi} > 0$ and $B_z > 0$, which is also a case where $H_j > 0$. Here the sign of $C_{0}$ is reversed. This field has the same signs as Model 2 in Fig.~\ref{fig:simple-helix}.

In the Southern hemisphere, the dipolar and quadrupolar models are the same, with $B_\phi > 0$ and $B_z < 0$.  The signs here are the same as Model 4, but since Model 4S is the same as Model 3N, we should compare Model 3N in Fig.~\ref{fig:freq-dependence} to the right plot in Fig.~\ref{fig:freq-dependence-dynamo}, we also find many similarities between the two plots, with the same differences for $B_{\text{rms}}$ as noted in the Northern hemisphere.

An important point to notice is that, for the dipolar dynamo field, the cross-correlation coefficient has opposite signs in the two hemispheres. In the quadrupolar dynamo field, the cross-correlation coefficient has same signs in the two hemispheres.

\subsection{Summary of the models}
\label{sec:modeldiscussion}

From the tests that we performed with the simple helical field (Fig.~\ref{fig:freq-dependence}), we are able to make the following statements:
\begin{enumerate}
\item Intrinsically, a helical field looks like a Faraday rotated one, and there are parts of the observed emission pattern that are depolarized due to the geometry of the field. Thus, large-scale helicity does introduce an intrinsic correlation between $\text{FD}$ and polarized fraction, even in instances where there is negligible Faraday rotation.

\item The sign of the cross-correlation coefficient does not, in general, correspond to the sign of the helicity. However, measurements of $C$ as a function of $\nu$ show trends that can help distinguish the cases presented in Table~\ref{tab:helix-cases}. In addition, in the case of a simple helical field, if we know the sign of $B_{\phi}$ from other measurements, combined with the sign of $C_0$, we can infer the sign of the helicity. 

\item Introducing a smaller-scale random magnetic field, expected in a turbulent medium, has a significant impact on this picture. A significant result is that the ratio between the coherent and random components is an important indicator as to whether $|C|$ can be useful in the detection of the helicity in the large-scale coherent component. If the random component has a magnitude that is of the same order as the helical coherent component, then we expect $|C|\ne0$. However, if the random component is much larger than the coherent component, then we find $C$ close to zero.
\end{enumerate}

We expect that the more complex and physically motivated case using the dynamo model should show properties of a field with helicity, since this is a predicted consequence of dynamo theory. Indeed the frequency dependence of $C$ in the dynamo model, shown in Figs.~\ref{fig:freq-dependence-dynamo} and \ref{fig:freq-dependence-dynamo-quad}, has similar trends as the simpler case in Fig.~\ref{fig:freq-dependence}. By comparing the two cases,  we can convincingly say that the 
Northern hemisphere of the dynamo model is consistent with right-handed helicity and the Southern hemisphere is consistent with left-handed helicity in both the dipolar and quadrupolar cases. 
The trends in the plots showing $C$ as a function of $\nu$ for this more complex model are largely consistent with those in the simple case. This gives us further confidence that these trends can indicate helicity in the large-scale magnetic field.

In addition to the sign of $C_0$, which is where Faraday rotation is negligible, we also note that we may be informed by the slope of $C(\lambda)$, which describes how $C$ changes as the amount of Faraday rotation (i.e., as wavelength) increases from $\lambda=0$~m ($\nu=\infty$) to $\lambda\approx0.2$~m ($\nu\approx1.4$~GHz). In the case of right-handed helicity ($H_j>0$), $C$ gets smaller as the amount of Faraday rotation increases. This is seen for both Models 1N and 2N in Fig.~\ref{fig:freq-dependence}. In the case of left-handed helicity, $C$ gets larger as the amount of Faraday rotation increases, which is seen for both Models 3N and 4N in Fig.~\ref{fig:freq-dependence}. This trend is also true in the dynamo models (see Figs.~\ref{fig:freq-dependence-dynamo} and \ref{fig:freq-dependence-dynamo-quad}) in the cases where $B_{\text{rms}}=0~\mu$G and $B_{\text{rms}}=1~\mu$G. However, for $B_{\text{rms}}=3~\mu$G and $B_{\text{rms}}=6~\mu$G in the dynamo model, we see the opposite trend. It is not clear if the trend seen in all cases for the $B_{\text{rms}}=0~\mu$G case is true in general, for all instances of a helical field, and for observers in any location. This requires further study.

The trend we observe can be directly compared to the result of \citet{Volegova:2010go} (see their Fig.~5) and \citet{Brandenburg:2014gp}. They find that for $H_j>0$, $C$ gets larger as the amount of Faraday rotation increases, which is opposite to what we find in the $B_{\text{rms}}=0~\mu$G case. The likely reason for this is that we have taken Faraday rotation to be right-handed about the magnetic field \citep[e.g.,][]{2018arXiv180607391R}, whereas their assumptions imply that it is left-handed (Brandenburg \& Stepanov, private communication). When we repeat our experiment with left-handed Faraday rotation, we find that our results do agree with theirs. In addition, although we both use helical fields, the geometry of the two scenarios are quite different. We use a large-scale, coiled field (like a slinky), whereas \citet{Volegova:2010go} and \citet{Brandenburg:2014gp} use a ``staircase'' type helical field \citep[see Fig.~8 in][]{Brandenburg:2014gp}. It is not clear whether we should expect that these two cases should give the same result for the trend of $C(\lambda)$. Of the two, the slinky-type geometry used in this work is more consistent with that which is typically used in large-scale models of the Galactic magnetic field \citep[e.g.,][]{Collaboration:2016eh, Jansson:2012ep}.

\section{Discussion}
\label{sec:discussion}

Although the precise value is not known, we expect the magnitude of the random magnetic field component in the Milky Way Galaxy to be $B_{\text{rms}}\sim6\mu$G, which is around two times larger than the coherent component \citep{2015ASSL..407..483H} (see discussion in Sec.~\ref{sec:modelling}). Thus we focus on comparing our observations to the case where $B_{\text{rms}}=3~\mu$G in Fig.~\ref{fig:freq-dependence} and $B_{\text{rms}}=6~\mu$G in Figs.~\ref{fig:freq-dependence-dynamo} and \ref{fig:freq-dependence-dynamo-quad}, which are the cases where the random component is around two times larger than the coherent component. 

The rotation of the stellar component of the Milky Way Galaxy is known to be clockwise, as observed from above the North Galactic Pole \citep[e.g.,][]{1927BAN.....3..275O}. Thus, according to the model presented by \citet{1970ApJ...162..665P}, we would expect $H_j < 0$ for the Northern hemisphere and $H_j > 0$ for the Southern hemisphere, i.e., left-handed about the z-axis. Given that the measurements for real data show that $C_0>0$ in both hemispheres, this would be consistent with a quadrupolar case where $B_z$ and $H_j$ both change sign, and $C_0>0$ in both hemispheres. Thus we find this is consistent with Model 3N in the Northern hemisphere, where $H_j < 0$ and $C_{0}>0$, and Model 1S in the Southern hemisphere, where $H_j > 0$ and $C_{0}>0$ (recalling that Model 1S = Model 2N).

In Fig.~\ref{fig:freq-dependence} we can see that for the $B_{\text{rms}}=3~\mu$G case, Model 2N (i.e., Model 1S) has a value of $C_{0}$ that is consistent with both of our Southern hemisphere measurements. In the $B_{\text{rms}}=3~\mu$G case of Model 3N, the value of $C_{0}$ is somewhat less than our Northern hemisphere WMAP measurement, though it agrees with the Northern hemisphere Planck measurement. This agrees with the Parker model predictions from the stellar rotation direction. This scenario, i.e., left-handed helicity in the Northern hemisphere and right-handed helicity in the Southern hemisphere, with a $B_z$ pointing away from the observer in both cases, is most consistent with what we detect in the observations when compared to the simple helix model.

However, in the quadrupolar case of the dynamo model, shown in Fig.~\ref{fig:freq-dependence-dynamo-quad}, we find the opposite. We find that the WMAP and Planck measurements in the Northern hemisphere agree best with the Northern hemisphere of this dynamo model (i.e., which is similar to Model 2N of the simple helix) and that the WMAP and Planck measurements in the Southern hemisphere agree best with the Southern hemisphere of this dynamo model (i.e., which is similar to Model 3N of the simple helix). This scenario, i.e., right-handed helicity in the Northern hemisphere and left-handed helicity in the Southern hemisphere, with a $B_z$ pointing away from the observer in both cases, is most consistent with what we detect in the observations when compared to the dynamo model.

The models to which we are comparing are very simple and surely do not capture a complete picture of the Milky Way field. We note that observations of the Milky Way Galaxy reveal the presence of magnetic reversals in its disk field \citep[e.g.,][]{2007ApJ...663..258B}. Further, recent observations also reveal the presence of magnetic reversals in the halo fields of some external galaxies \citep[e.g.,][]{2019A&A...632A..11M,2020A&A...639A.112K}. These reversals may help explain the ambiguity that we find in the direction of the toroidal component of the field (i.e., when comparing the simple helical field with the dynamo fields).

In both cases we find that $B_z$ should point away from the observer in both hemispheres. This disagrees with the result of \citet{2010ApJ...714.1170M}, which found no evidence for a $B_z$-component in the Northern hemisphere. They found a positive average rotation measure in the Southern hemisphere, which is evidence that the $B_z$-component points towards the observer, and not away. As pointed out in the case of the toroidal component, the models to which we compare are very simple, which means a more complete, and complex, model is likely needed to explain these discrepancies.

We can make the following observations:
\begin{enumerate}

\item The correlation we find in the Planck and WMAP data cannot be due to helicity in the random component of the magnetic field, since that requires Faraday rotation, and the frequency of these observations is too high for Faraday rotation to be a significant factor.

\item Any explanation for the observed correlation requires a coherent pattern in the magnetic field over large angular scales. 

\item An analysis of how the cross-correlation coefficient in observations changes as a function of frequency would provide an additional diagnostic.
\end{enumerate}

Helicity in the large-scale field is one possible - and plausible - explanation for the correlations that we find, however, it is not the only explanation. The local magnetic field, and possibly helicity within the local field itself, can not be ruled out as a possible cause of these correlations. It is reasonable to expect the local magnetic field to have non-zero helicity, because under the model of \citet{1970ApJ...162..665P}, the expansion of the Local Bubble was presumably accompanied by counterrotation with respect to Galactic rotation, under the effect of the Coriolis force.

The North Polar Spur (NPS) is the most dominant feature in the Northern Galactic hemisphere and has the highest polarized fraction. The origin of the NPS, and related spur-like features, are unknown but are most likely local features \citep[][]{2015ApJ...811...40S,2020West}. Regardless of their origin, the presence of the NPS and related Northern hemisphere loops will impact the results in the Northern Galactic hemisphere and may contribute to the discrepancy and lack of correlation observed in the Planck data. Other, fainter, loops may impact the result in the Southern hemisphere, but to a lesser degree. We consider masking out these features, however the full extent of their emission is not well understood, so it is unclear how to do this precisely. Experimenting with different masks is beyond the scope of this paper. In addition, a very few extragalactic sources that could potentially impact the results are visible in the maps (e.g., Centaurus~A, the Large and Small Magellanic clouds). Due to their relatively small angular extent, the bootstrapping method we use to calculate the correlation (see Sec.~\ref{sec:data}) should mitigate any possible impact these might have on the value of $C$. 

In our models, we removed the contribution of local features from the thermal electron density model of \citet{2017ApJ...835...29Y}. Being nearby, these features will have a large angular extent on the sky and will have a significant impact on the measured FD across the sky. The significance that these features might have on these correlations should be investigated in future work.

We also expect helicity in the large-scale field of opposite sign to the very small-scale field \citep[e.g.,][Fig.~9.6]{2005PhR...417....1B}, which we have not included here. However, given point (i) earlier in this section, and given that these very small-scale effects (i.e., much smaller than the injection scale of turbulence) would be difficult to detect with the low resolution of our data and models, we believe it is reasonable to neglect the very small-scale component.

\section{Conclusions}
\label{sec:conclusions}

We find that a cross-correlation between FD and polarized fraction of the synchrotron radio emission from the Milky Way Galaxy has a positive and measurable value in the Southern Galactic hemisphere, as shown in Table~\ref{tab:correlations}. For the Northern hemisphere, we find a smaller, but still measurably positive value of the cross-correlation coefficient, $C$, in WMAP data, but the value in Planck data is consistent with zero. We conclude that our measurements of $C$ are consistent with the presence of helicity in the mean magnetic field of the Galaxy.

We demonstrate that a model of large-scale magnetic field can exhibit a correlation that is similar to the result for helical turbulence shown by \citet{Volegova:2010go} and \citet{Brandenburg:2014gp}. From modelling we find that $C$, or at least the limit of $C$ for high frequencies, which we call $C_0$, has the same parity as the large-scale magnetic field: if the field is dipolar, $C_0$ changes sign across the midplane; if the field is quadrupolar, $C_0$ keeps the same sign. 

We have shown that the simulated synchrotron emission and Faraday rotation of a large-scale helical magnetic field have a complex relationship with $C$ that varies as a function of frequency (Figs.~\ref{fig:freq-dependence},  \ref{fig:freq-dependence-dynamo}, and \ref{fig:freq-dependence-dynamo-quad}). A follow-up study of multi-frequency observations would help to confirm these results. We plan to present an analysis of lower frequency ($\sim1$~GHz) data from the Global Magneto-Ionic Medium Survey \citep[GMIMS,][]{2009IAUS..259...89W} in a forthcoming study.

Comparing our simple helical magnetic field models with the observations, we find the observational measurements are consistent with the presence of a quadrupolar magnetic field with left-handed helicity in the Northern hemisphere and right-handed helicity in the Southern hemisphere, with a vertical component of the magnetic field that is pointing away from the observer in both hemispheres (similar to Model 3 in the North and Model 2 in the South, see Fig.~\ref{fig:freq-dependence}). The comparison of our dynamo model with the observations is also consistent with a quadrupolar magnetic field with a vertical component of the magnetic field that is pointing away from the observer in both hemispheres, however in this case we find it is more consistent with right-handed helicity in the North and left-handed helicity in the South. 

Further work is needed to resolve this discrepancy, and demonstrate that this is a global feature of the large-scale field rather than a local one. The study presented in this work is limited by angular resolution and may benefit from careful masking of local features. Broadband FD observations offered by upcoming surveys such as the Polarization Sky Survey of the Universe's Magnetism (POSSUM) \citep{2010AAS...21547013G}, which uses the Australian Square Kilometer Array Pathfinder (ASKAP) telescope and the Very Large Array Sky Survey (VLASS) \citep{Lacy_2020} are expected to provide ten times the source density of current observations. These observations will provide a much improved Galactic FD map and will help verify the results of this work. However, even upcoming surveys, with greatly improved resolution, will not be able to access the very small scales at which helicity of opposite sign is expected.

 Improved Galactic magnetic field modelling is also necessary to verify these results. The IMAGINE consortium and the Bayseian inference code they are developing \citep{haverkorn2019imagine} aims to develop more sophisticated Galactic magnetic field modelling using dynamo models such as those in GALMAG \citep{2019A&A...623A.113S}, which include higher-order dynamo modes and combinations of modes. The cross-correlation analysis presented in this work should be applied to an improved model for the Galactic magnetic field when such a model becomes available. 

However, even if the phenomenon is local, and even if the sign of the helicity in each hemisphere is only suggestive rather than conclusive, these results strongly indicate a detection of non-zero helicity. This strengthens the argument that ordered fields on galactic scales arise through dynamo action.

\section*{Acknowledgements}

The Dunlap Institute is funded through an endowment established by the David Dunlap family and the University of Toronto. J.L.W. and B.M.G. acknowledge the support of the Natural Sciences and Engineering Research Council of Canada (NSERC) through grant RGPIN-2015-05948, and of the Canada Research Chairs program. We thank Niels Oppermann for useful input during the early parts of this project. We thank A. Brandenburg and R. Stepanov for helpful discussions that improved this work. We also thank the anonymous referee for their careful comments, which greatly improved this manuscript. This research has made use of the NASA Astrophysics Data System (ADS), Maple 2018 (Maplesoft, a division of Waterloo Maple Inc., Waterloo, Ontario), and the following python packages: astropy \citep{Astropy}, healpy \citep{Healpy}, matplotlib \citep{Matplotlib}, numpy \citep{Numpy}, pandas \citep{Pandas}, scipy \citep{Scipy}, and seaborn \citep{seaborn}. 

\section*{Data Availability}
The data used in this study are all publicly available. The Planck foreground maps can be downloaded from: \url{https://pla.esac.esa.int/}. Similarly, the WMAP data is available at \url{https://lambda.gsfc.nasa.gov/product/map/current/}. The Galactic Faraday depth map from \citet{2015A&A...575A.118O}, is available at \url{https://wwwmpa.mpa-garching.mpg.de/ift/faraday/2014/index.html}.




\bibliographystyle{mnras}
\bibliography{references} 




\appendix

\section{Dynamo models}
\label{sec:dynamoappendix} 
The dynamo models that we use start from the classical mean-field dynamo equations \citep{1978mfge.book.....M}
\begin{equation}
{\partial _t}{\bf{A}} = {\bf{v}} \times \nabla  \times {\bf{A}} - \eta \nabla  \times \nabla  \times {\bf{A}} + {\alpha _d}\nabla  \times {\bf{A}},
\end{equation}
where $\bf{A}$ is the large-scale (mean) vector potential of the mean magnetic field ${\bf{B}}$ (and $\bf{B}  = \nabla  \times \bf{A}$). Here, $\bf{v}$ is the mean velocity with components $(u,v,w)$, $\eta$ is the resistive diffusivity, and ${\alpha _d}$ is an effective ``twisting'' velocity that describes the alpha effect, which leads to the macroscopic, large-scale magnetic helicity, $H_m$ (see Sec.~\ref{sec:intro}).

The model is parameterized using the variables $R$, $\Phi$, $Z$, $a$, $u$, $v$, $w$, $T$, $m$, $q$, $\epsilon$, $C1$, and $C2$.

The cylindrical coordinates $\{r, \phi,z\}$ are transformed into scale invariant coordinates $\{R, \Phi, Z\}$ according to: 
\begin{equation}
r = R e^{\delta T} , \Phi = \phi + (\epsilon+q) \delta T, z = Z e^{\delta T} 
\end{equation}
\citep[e.g.][]{henriksen2015scale}, where $\delta$ is an arbitrary scale that appears in the spatial scaling, $\epsilon$ is a number that fixes the rate of rotation of the magnetic field in time, and $T$ is a time variable that follows $e^{\alpha T} = 1 + \widetilde{\alpha}_d \alpha t$ ($\widetilde{\alpha}_d$ is a numerical constant that appears in the scale invariant form for the helicity, $\alpha$ is an arbitrary scale used in the temporal scaling). The variable $a \equiv \alpha / \delta $ is a parameter of the model defined as the self-similar `class' \citep{1991JMP....32.2580C}, which reflects the dimensions of a global constant. 

The time dependence of the models does not affect the field geometry because the dependence is mainly a multiplicative power law or exponential in time, depending on the parameter $a$. The one exception is the rotation, which can change the position of the observer relative to the structure of the field. This can be dealt with by varying the parameter $\epsilon$.

Further assumptions allow the model solutions to be simplified further into two cases: one where outflow and accretion are allowed to vary but the rotation is held constant and another where rotation in the ``pattern frame'' is allowed  \citep[i.e., the rest frame of the dynamo magnetic field, see][]{Henriksen:2017gp}. In the rotation-only case, $u=w=0$, but $v$ and $a$ are allowed to vary. Whereas in the outflow case, $a=1$, $u=v=0$ and $w$ is allowed to vary (where $w>0$ is outflow and $w<0$ is inflow).

The pitch angle, $\psi$, of the spiral is set by the variable $q$, where $1/q$ ($q$ positive) is the tangent of the  angle that a trailing (assuming the $\phi$ direction is in the direction of galactic rotation) spiral arm makes with the circular direction \citep[][see equation 9]{Henriksen:2018uk} and $\psi=\arctan(1/q)$. We use $q=4.9$, which corresponds to $\psi=-11.5^{\circ}$, the value used in most Milky Way magnetic field models \citep{Collaboration:2016eh}.

The `spiral mode', $m$, arises in the spirally symmetric case when solving for the magnetic field potential $A$. Solutions are searched
for in the complex form 
\begin{equation}
\begin{aligned}
& A (R, \Phi, Z) = A (\zeta)e^{im \kappa}
\\
& \zeta=Z/R
\\
& \kappa \equiv \Phi + q \ln R \equiv \phi + q \ln(r) + \epsilon T.
\end{aligned}
\end{equation}

In these models, the $m=0$ mode has the diverging X-shaped morphology that is similar to what is observed in external galaxies. All modes have a spiral morphology of varying degree. 
\citet{2019MNRAS.487.1498W} use a mixture of these modes to attempt to model the observed magnetic field of NGC~4631.

Finally, $C1$ and $C2$ are the boundary conditions. The boundary condition at the disc must be treated carefully so that the solutions are continuous across the disk along the real axis  \citep[][see Sec. 4.1]{Henriksen:2018uk}.

\begin{figure}
\centering 
\begin{minipage}{4cm}
\includegraphics[width=4cm]{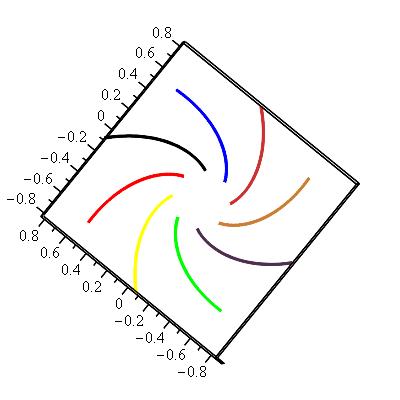}
\end{minipage}
\hfill
\begin{minipage}{4cm}
\includegraphics[width=4cm]{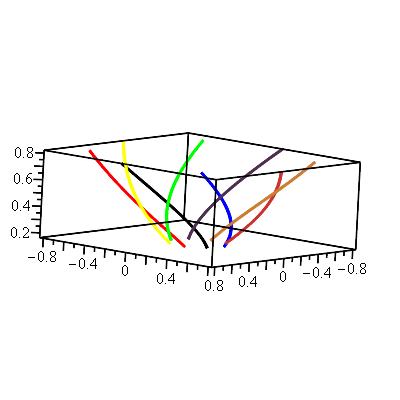}
\end{minipage}

\begin{scriptsize}\caption{\label{fig:3Dfieldlines}3D view of a selection of magnetic field lines for the dynamo case described as seen from the top down (left) and from the side for $z>0$ (right).}
\end{scriptsize} 

\end{figure}

For this work, we use a case with $m=0$, $a=1$, $(u,v,w) =  (0, 0, 1)$, $T=1$, $q=4.9$, $\epsilon=-1$, and $(C1, C2) =  (0, 1)$. In Fig.~\ref{fig:3Dfieldlines} we show a 3D plot of this case for $z>0$. 

\bsp	
\label{lastpage}
\end{document}